\documentclass[superscriptaddress,showpacs,amsmath,amssymb,twocolumn,nofootinbib]{revtex4-1}

\usepackage{amsmath, amsthm, amssymb, bm, braket}
\usepackage[dvips]{graphicx}
\usepackage{color}
\usepackage[normalem]{ulem}

\begin{document}

\title{Magnetic dipole excitation and its sum rule in nuclei with two valence nucleons}

\author{Tomohiro Oishi}
\email[E-mail: ]{toishi@phy.hr} 
\author{Nils Paar}
\email[E-mail: ]{npaar@phy.hr}
\affiliation{Department of Physics, Faculty of Science, University of Zagreb, Bijeni\v{c}ka c. 32, 10000 Zagreb, Croatia}

\renewcommand{\figurename}{FIG.}
\renewcommand{\tablename}{TABLE}

\newcommand{\bi}[1]{\ensuremath{\boldsymbol{#1}}}
\newcommand{\unit}[1]{\ensuremath{\mathrm{#1}}}
\newcommand{\oprt}[1]{\ensuremath{\hat{\mathcal{#1}}}}
\newcommand{\abs}[1]{\ensuremath{\left| #1 \right|}}

\newcommand{\crc}[1] {c^{\dagger}_{#1}}
\newcommand{\anc}[1] {c_{#1}}
\newcommand{\crb}[1] {a^{\dagger}_{#1}}
\newcommand{\anb}[1] {a_{#1}}

\newcommand{\slashed}[1] {\not\!{#1}} 

\def \beq{\begin{equation}}
\def \eeq{\end{equation}}
\def \beqa{\begin{eqnarray}}
\def \eeqa{\end{eqnarray}}

\def \bir{\bi{r}}
\def \ubir{\bar{\bi{r}}}
\def \bip{\bi{p}}
\def \ubip{\bar{\bi{r}}}

\def \adel{\tilde{l}} 

\begin{abstract}
{\noindent
{\bf Background:} 
Magnetic dipole (M1) excitation is the leading mode of nuclear 
excitation by the magnetic field, which couples unnatural-parity states. 
Since the M1 excitation occurs mainly for open-shell nuclei, 
the nuclear pairing effect is expected to play a role. 
As expected from the form of operator, this mode may 
provide the information on the spin-related properties, 
including the spin component of dineutron and diproton correlations. 
In general, the sum rule for M1 transition strength has not been derived yet. \\
{\bf Purpose:} 
To investigate the M1 excitation of the systems with two valence nucleons 
above the closed-shell core, with pairing correlation included, and 
to establish the M1 sum rule that could be used to validate theoretical and 
experimental approaches. 
Possibility to utilize the M1 excitation as a tool to investigate 
the pairing correlation in medium is also discussed. \\
{\bf Method:} 
Three-body model, which consists of a rigid spherical core and two valence 
nucleons, is employed. 
Interactions for its two-body subsystems are phenomenologically 
determined in order to reproduce the two-body and three-body energies. 
We also derive the M1 sum rule within this three-body picture. \\
{\bf Conclusion:} 
The introduced M1 sum rule can be utilized as a benchmark for model
calculations of M1 transitions in the systems with two valence nucleons. 
The total sum of the M1 transition strength is related with 
the coupled spin of valence nucleons in the open shell, where the 
pairing correlation is unnegligible. 
The three-body-model calculations for $^{18}$O, $^{18}$Ne, and $^{42}$Ca nuclei 
demonstrate a significant effect of the pairing correlations on the low-lying M1 transitions. 
Therefore, further experimental studies of M1 transitions in those systems 
are on demand, in order to validate proposed sum rule, provide a suitable probe for 
the nuclear pairing in medium, as well as to optimize the pairing models. 
}
\end{abstract}

\maketitle

\section{Introduction} \label{Sec:intro}
Electromagnetic excitations in finite nuclei are fundamental 
in nuclear physics and astrophysics. 
They provide valuable probe of the nuclear structure 
and dynamics and play a decisive role in the processes in stellar environments. 
In particular, magnetic dipole (M1) 
transitions have invoked various 
interests and discussions in recent studies \cite{1985Rich,1996Kneis_Exp_Rev,2008Pietralla_Rev,2010Heyde_M1_Rev}. 
The M1 excitation mode is relevant to a diversity 
of nuclear properties, including 
unnatural-parity states, 
spin-orbit splitting, 
tensor force effect, etc. 
Several collective nuclear phenomena, including 
scissors mode in deformed nuclei, can be activated by the 
M1 excitation \cite{1984Bohle,1984Bohle_02,1990Rich,1990Taka,1996Kamerdz,2009Vesely,2010Nest,2010Nest_2,2017Schwegner_M1,2018Balb}. 
In addition, the correspondence between the M1 
and the zeroth component of Gamow-Teller modes 
has been discussed \cite{2009Vesely,2004Lang,2007Stone}: 
the M1 excitation can be utilized to optimize theoretical methods 
to predict the Gamow-Teller resonance distribution, which plays an essential role 
in neutrino-nucleus scattering.

Within a shell-model picture, the M1 transition couples spin-orbit partner orbits. 
Thus, this process is measurable mainly in open-shell nuclei, 
where the transition from the lower to higher partner orbit is 
available \cite{1970Moreh_M1_Exp,1985Lasz_M1_Exp,1988Lasz_M1_Exp,1990Rich,1993Kam,1984Bohle,1984Bohle_02,2003Fearick,2016Pai,2017Shizuma}. 
In such systems, nuclear pairing correlation can 
play an essential role \cite{05BB,13BZ,03Dean_rev,03Bender_rev}. 
However, compared with the electric dipole (E1) 
and quadrupole (E2) modes \cite{2005HS,2009Yoshida_pyg,2011TE,2011Stetcu,2014Ebata_pyg,2013Dang,2011Oishi}, 
the knowledge on the pairing effect on magnetic modes, 
as well as on unnatural-parity states, is 
rather limited \cite{1993Kam,2009Vesely,2010Nest,2010Nest_2}. 



In studies of nuclear modes of excitation, the sum rules associated 
to the transition strength and energy-weighted sum rules, represent 
an essential tool for the analyses of the excitations, not only as benchmark 
tests of the theoretical frameworks involved, but also to inspect the completeness of the experimental 
data~\cite{70Eisenberg,80Ring,1999Wang,2018Lu,1963Ikeda,1964Fujita,2015Johnson,2015Hino,2018Bonasera,2018Garg,2018Gupta,2016Sagawa,2016Gambacurta,2009Mason}. 
Over the past decades, the analyses of the sum rules have been 
mandatory to validate theoretical approaches to describe various 
modes of excitation. In particular, the Fermi sum rule and Gamow-Teller sum 
rules are well established for the isospin-flip and spin-flip isospin-flip 
charge-exchange resonances, respectively~\cite{1963Ikeda,1964Fujita,1992Osterfeld}. 
The sum rules for electric multipole modes have also been established 
and very useful in studies of giant monopole, dipole and quadrupole 
resonances~\cite{70Eisenberg,80Ring,1999Wang}. 
Recently, the general formulas have been derived for non-energy-weighted and 
energy-weighted sum rules of electric and weak transitions within 
a shell model occupation-space framework~\cite{2018Lu}.
However, so far, the general version of sum rule has not been 
established for the M1 excitation. 
It represents an open problem of relevance for complete understanding of magnetic transitions.

In this work, we explore the properties of M1 excitations with 
a strong pairing, by employing the three-body model. 
This model is applicable to nuclei 
composed of the closed-shell core and two valence nucleons. 
In this model assumption, the sum rule of M1 transition is also introduced. 
Its implementation for the purpose of the present study may provide 
a standard to benchmark the theoretical 
investigation to some nuclei, where the three-body picture 
can be a good approximation. 



It is worthwhile to mention so-called dineutron and diproton correlations, 
which can be relevant to our interest. 
Several theoretical studies have predicted that, 
under the effect of the pairing correlation in medium, 
the valence two-nucleon subsystem may 
have a spin-singlet ($S_{12}=0$) and/or spatially localized 
structure \cite{05Mats,12Mats,2005HS,07Bertulani_76,07Hagi_01,14Hagi_2n,2010Oishi,2014Oishi}. 
However, considering its measurability, there remain some problems 
to clarify this intrinsic structure \cite{2016Hagino,2016Kiku,2018Gri}. 
As expected from the form of the M1 transition operator, 
it may provide a suitable way to probe 
the coupled spin of nucleons, $S_{12}$, 
which should reflect the pairing effect inside nuclei. 

Our theoretical formalism, as well as numerical methods, 
are described in section \ref{Sec:3BM}. 
Then, section \ref{Sec:Result} is devoted to present 
our results and discussion. 
Finally, the summary and conclusion of this work are given in section \ref{Sec:Sum}. 
The CGS-Gauss system of units is used in this article.

\section{Three-body model} \label{Sec:3BM}
We employ the three-body model, which has been developed in 
Refs. \cite{2005HS,07Hagi_01,07Bertulani_76,14Hagi_2n,88Suzuki_COSM,1991BE,1997EBH,2010Oishi,2014Oishi,2014Lorenzo}. 
Namely, the system consists of a rigid-core nucleus and two valence 
nucleons with the following assumptions. 
\begin{itemize}
  \item Core nucleus is spherical, i.e., with the shell-closure and $J^{\pi} = 0^+$. Nucleons in the core are not active for excitation. 
  \item Two valence nucleons are of the same kind, namely, protons or neutrons. They feel the mean field, $V_C$, generated by the core nucleus. 
\end{itemize}
Thus, the three-body Hamiltonian reads 
\beqa
 H &=& h_C(1) + h_C(2) + v_{NN}(\bir_1,\bir_2) + x_{\rm rec}, \label{eq:3B_Ham} \\
 && h_C(i) = \frac{p^2_i}{2\mu_i} + V_{C}(\bir_i), \label{eq:SP_Ham} \\
 && x_{\rm rec} = \frac{\bip_1 \cdot \bip_2}{m_C}~~~{\rm (recoil~term)}, \label{eq:Recoil}
\eeqa
where $i=1$ ($2$) indicates the first (second) valence nucleon. 
The single-particle (SP) Hamiltonian, $h_C$, contains 
the mass parameters, $\mu_i=m_N m_C /(m_N + m_C)$, 
$m_N=939.565$ ($938.272$) MeV$/c^2$ for neutrons (protons), and 
$m_C$ of the core nucleus. 
Note that the recoil term also exists after the center-of-mass 
motion is subtracted. 
The SP potential, $V_C$, and the pairing potential, $v_{NN}$, 
are determined in the next section.

The SP state, $\psi_{nljm}(\bir)=R_{nlj}(r) \cdot \mathcal{Y}_{ljm}(\theta,\phi)$, 
is solved to satisfy 
\beq
 h_C(i) \psi_{n_i l_i j_i m_i}(\bir_i) = e_i \psi_{n_i l_i j_i m_i}(\bir_i). 
\eeq
We employ the SP states up to the $h_{11/2}$ channel. 
In order to take into account the Pauli principle, 
we exclude the states occupied by the core. 
Continuum states are discretized within a box, $R_{\rm box}=30$ fm. 
The cutoff energy, $E_{\rm cut}=30$ MeV, is also employed 
to truncate the model space. 
We checked that this truncation indeed provides a 
sufficient convergence for results in the following sections.

For diagonalization of the total Hamiltonian, we employ 
the anti-symmetrized two-particle (TP) states for basis. 
That is, 
$\tilde{\Psi}^{(J,M)}_{k_1 k_2} (\bir_1, \bir_2) \equiv \oprt{A} \Psi^{(J,M)}_{k_1 k_2} (\bir_1, \bir_2)$, 
where 
\beqa
 && \Psi^{(J,M)}_{k_1 k_2} (\bir_1, \bir_2) \nonumber \\
 && \equiv \sum_{m_1 m_2} \mathcal{C}^{(J,M)j_1 j_2}_{m_1 m_2} \psi_{n_1 l_1 j_1 m_1}(\bir_1) \psi_{n_2 l_2 j_2 m_2}(\bir_2).
\eeqa
Here we take the short-hand label, $k_i \equiv \left\{ n_i l_i j_i \right\}$. 

The M1 operator, which operates only on the 
two valence nucleons, reads 
\beq
 \oprt{Q}_{\nu} = \hat{q}_{\nu}(1) + \hat{q}_{\nu}(2),
\eeq
where
\beqa
 \hat{q}_{\nu=0} &=& \mu_{\rm N} \sqrt{\frac{3}{4\pi}} (g_l \hat{l}_0 + g_s \hat{s}_0), \nonumber \\
 \hat{q}_{\nu=\pm 1} &=& (\mp) \mu_{\rm N} \sqrt{\frac{3}{4\pi}} (g_l \hat{l}_{\pm} + g_s \hat{s}_{\pm}). 
\eeqa
Notice that $\hat{l}_0=\hat{l}_z$, $\hat{l}_{\pm}=(\hat{l}_x \pm i \hat{l}_y)/\sqrt{2}$, 
and similarly for the spin operators. 
As well known, $g$ factors are given as $g_l=1~(0)$, 
and $g_s=5.586~(-3.826)$ for the proton (neutron) \cite{70Eisenberg,80Ring}. 
In the following, we omit the nuclear magneton $\mu_N$ 
and the factor $\sqrt{3/4 \pi}$, except when it needs.

In this work, we discuss up to the one-body-operator level of the M1 excitation, 
whereas the meson-exchange-current effect is not included. 
In some ab-initio calculations, it has been shown that this effect indeed 
can contribute in addition to the M1 transition of the one-body-operator 
level \cite{1990Richter,2008Marcucci,1994Moraghe}. 
For treatment of this effect, one needs to consider the relevant multi-body terms. 
This is, however, technically demanding, and beyond the scope of this work. 
We notify that evaluation of this meson-exchange-current effect 
is awaiting for the future progress.

\subsection{M1 sum rule}
The sum-rule value (SRV) of M1 transitions is determined as 
\beqa
 S_{\rm M1} &\equiv& \sum_{f} \left( \abs{\Braket{f| \oprt{Q}_0 |i}}^2 \right. \nonumber \\
 & & + \left. \abs{\Braket{f| \oprt{Q}_{+} |i}}^2 + \abs{\Braket{f| \oprt{Q}_{-} |i}}^2 \right), 
\eeqa
where $\ket{i}$ and $\ket{f}$ indicate the initial 
and final states, respectively.
Within the three-body model, because only two nucleons are available to excite, 
this can be reduced to 
\beqa
 S_{\rm M1} &=& \Braket{i| \left( g_l\hat{\bf L}_{12} + g_s\hat{\bf S}_{12} \right)^2 |i} \nonumber \\
 &=& (g^2_l -g_l g_s) \Braket{\hat{\bf L}^2_{12}}^{(i)} + (g^2_s -g_l g_s) \Braket{\hat{\bf S}^2_{12}}^{(i)} \nonumber \\
 & & + g_l g_s \Braket{\hat{\bf J}^2}^{(i)}, \label{eq:SRV_1}
\eeqa
where $\hat{\bf J}=\hat{\bf L}_{12}+\hat{\bf S}_{12}$, $\hat{\bf L}_{12}=\hat{\bf l}(1)+\hat{\bf l}(2)$, and 
$\hat{\bf S}_{12}=\hat{\bf s}(1)+\hat{\bf s}(2)$. 
Namely, the SRV contains the information on the coupled spins of the valence nucleons 
at the initial state. 
Considering the simplification of this M1 sum rule, 
there are two cases, which can be especially worth mentioning. 
\begin{itemize}
   \item For two neutrons with $g_l=0$, SRV is simply 
determined from the initial spin-triplet component, $N^{(i)}_{S_{12}=1}$: 
\beqa
 && S_{\rm M1}({\rm 2n}) = g^2_s \Braket{\hat{\bf S}^2_{12}}^{(i)} \nonumber \\
 && = g^2_s \sum_{S_{12}=0,1} S_{12}(S_{12}+1) N^{(i)}_{S_{12}} = 2 g^2_s N^{(i)}_{S_{12}=1}. \label{eq:SRV_2n}
\eeqa
Notice that this spin-triplet component itself stands 
independently of the final-state properties. 
   \item When two protons or neutrons are coupled to $J_i=0$ at the initial 
state $\ket{i}$, Eq. (\ref{eq:SRV_1}) 
can be remarkably simplified. 
In this case, in terms of the $LS$-coupling scheme, 
the allowed components include $(L_{12},S_{12})=(1,1)$ and $(0,0)$, only. 
Thus, 
\beq
 S_{\rm M1}(J_i=0) = 2(g_l -g_s)^2 N_{(L_{12}=1,S_{12}=1)}^{(i)},
\eeq
where we have utilized that the first-component, $N_{(L_{12}=1,S_{12}=1)}^{(i)}$, 
can be identical to $N_{L_{12}=1}^{(i)}$ as well as $N_{S_{12}=1}^{(i)}$. 
\end{itemize}
Notice that, in both cases, the M1 SRV is enhanced (suppressed) 
when the $S_{12}=1$ ($S_{12}=0$) component is dominant. 
Therefore, the maximum limit of SRV is $2(g_l -g_s)^2$ when $N_{S_{12}=1}^{(i)}=100$\%.

\subsection{no-pairing SRV}
It is worthwhile to mention the SRV in one special case, where 
the non-diagonal terms in the Hamiltonian can be neglected: 
$H\rightarrow H_{\rm NP} = h_C(1)+h_C(2)$. 
In this case, the SRV can be analytically obtained. 
Because of the diagonal Hamiltonian, 
the initial state with the total angular momentum $J_i$ 
should be solved as the single set of coupled SP states. 
That is, 
\beq
 \ket{i} = \left[ \ket{l_1 j_1} \otimes \ket{l_2 j_2} \right]^{(J_i)}, 
\eeq
where $l_k$ and $j_k$ indicate the angle-quantum numbers of the $k$th valence nucleon. 
Thus, Eq. (\ref{eq:SRV_1}) is represented as 
\beqa
 && S_{\rm M1,NP} = (g^2_l -g_l g_s) \sum_{L_{12}} L_{12}(L_{12}+1) N_{L_{12}}^{(l_1 j_1,l_2 j_2; J_i)} \nonumber \\
 && ~~~+(g^2_s -g_l g_s) 2N_{S_{12}=1}^{(l_1 j_1,l_2 j_2; J_i)} + g_l g_s J_i(J_i+1), \label{eq:SRV_ANA}
\eeqa
where $N_X^{(l_1 j_1,l_2 j_2; J_i)}$ indicates the contribution of the $X$ component 
in this initial state. 
The analytic derivation of $N_X^{(l_1 j_1,l_2 j_2; J_i)}$ for an 
arbitrary set of $(l_1, j_1, l_2, j_2, J_i)$ can be found in, 
e.g. textbook \cite{60Edm}. 
This no-pairing, analytic SRV can provide a solid 
standard to benchmark one's theoretical model, as well as 
its computational implementation.

\section{Numerical Result} \label{Sec:Result}
\subsection{$^{18}$O nucleus}
For numerical calculation, first we focus on the $^{18}$O nucleus. 
Its core, $^{16}$O, is suitable to the rigid-core assumption: 
the lowest $1^+$ level of $^{16}$O locates at 
$13.6$ MeV \cite{NNDCHP}, which is sufficiently higher than the energy 
region we consider in the following. 
Thus, the core excitation by the M1 can be well separated 
from that of the two valence neutrons, 
allowing us to use the three-body model.

\subsubsection{no-pairing case}
For simplicity, we first neglect the non-diagonal 
terms, $v_{NN}$ and $x_{\rm rec}$ in Eq. (\ref{eq:3B_Ham}), 
thus we deal with the no-pairing Hamiltonian, $H_{\rm NP}=h_C(1)+h_C(2)$. 

In the first step, we need to constrain the core-neutron potential, $V_C$, 
for the SP states by considering the core-neutron subsystem, $^{17}$O. 
For this purpose, we employ the Woods-Saxon (WS) potential: 
\beqa
 V_C(\bir) &=& V_0 f(r) + U_{ls} ({\bf l}\cdot{\bf s}) \frac{1}{r} \frac{df}{dr}, \nonumber \\
 f(r) &=& \frac{1}{1+e^{(r-R_0)/a_0}}, \label{eq:WS}
\eeqa
where $R_0=r_0\cdot A_C^{1/3}$, $A_C=16$, $r_0=1.25$ fm, $a_0=0.65$ fm, 
$V_0=-53.2$ MeV, and $U_{ls}=22.1$ MeV$\cdot$fm$^2$. 
These parameters reproduce well the empirical SP energies of 
$^{17}$O, as shown in Table \ref{table:17O}. 
Note that, due to the Pauli principle, we exclude from our basis
$1s_{1/2}$, $1p_{3/2}$, and $1p_{1/2}$ states, which are 
occupied by the core.

\begin{table}[b] \begin{center}
\caption{Single-neutron energies for $^{17}$O (in MeV). 
For the resonant $d_{3/2}$ level, its energy and width 
$\Gamma$ are obtained by evaluating the scattering phase-shift. 
} \label{table:17O}
  \catcode`? = \active \def?{\phantom{0}} 
  \begingroup \renewcommand{\arraystretch}{1.2}
  \begin{tabular*}{\hsize} { @{\extracolsep{\fill}} rccc } \hline \hline
                    &This work       &Exp. \cite{NNDCHP}  &Type  \\  \hline
  $e(1d_{5/2})$      &$-4.143$        &$-4.143$  &bound  \\
  $e(2s_{1/2})$      &$-3.275$        &$-3.272$  &bound  \\
  $e_{\rm r}(d_{3/2})$ &$+0.902$         &$+0.941$  &resonance   \\
                   &($\Gamma=0.102$)  &($\Gamma=0.096$) \\ \hline \hline
  \end{tabular*}
  \endgroup
  \catcode`? = 12 
\end{center} \end{table}

In the following, we move to the core plus two-neutron system,
$^{18}$O. In the no-pairing case, the GS can be trivially solved as 
\beq
  \Psi_{\rm GS}(\bir_1, \bir_2) = \oprt{A} \left[ \psi_{1d_{5/2}} (\bir_1) \otimes \psi_{1d_{5/2}} (\bir_2) \right]^{(J_i=0)}, 
\eeq
which satisfies $H_{\rm NP} \ket{\Psi_{\rm GS}} =E_{\rm GS} \ket{\Psi_{\rm GS}}$ with 
$E_{\rm GS}=2e(1d_{5/2})$. 
Notice that, from Table \ref{table:17O}, this GS energy is 
$2e(1d_{5/2})=-8.286$ MeV, which is higher than the empirical 
two-neutron binding energy, $B_{\rm 2n}=-12.188$ MeV, of $^{18}$O \cite{NNDCHP}. 
This discrepancy is, of course, due to the no-pairing assumption. 

We now compute the M1 excitation from the $0^+$ ground state (GS). 
For $1^+$ excited states, we solve all the TP states coupled to $1^+$: 
$H_{\rm NP} \ket{f(1^+)} =E_f \ket{f(1^+)}$.

\begin{figure}[tb] \begin{center}
  \includegraphics[width = 0.95\hsize]{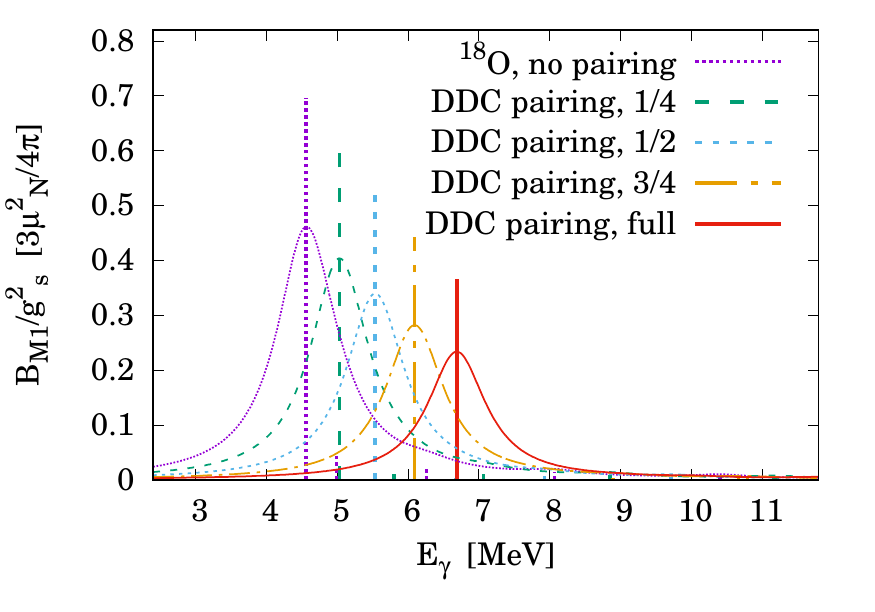}
  \includegraphics[width = 0.95\hsize]{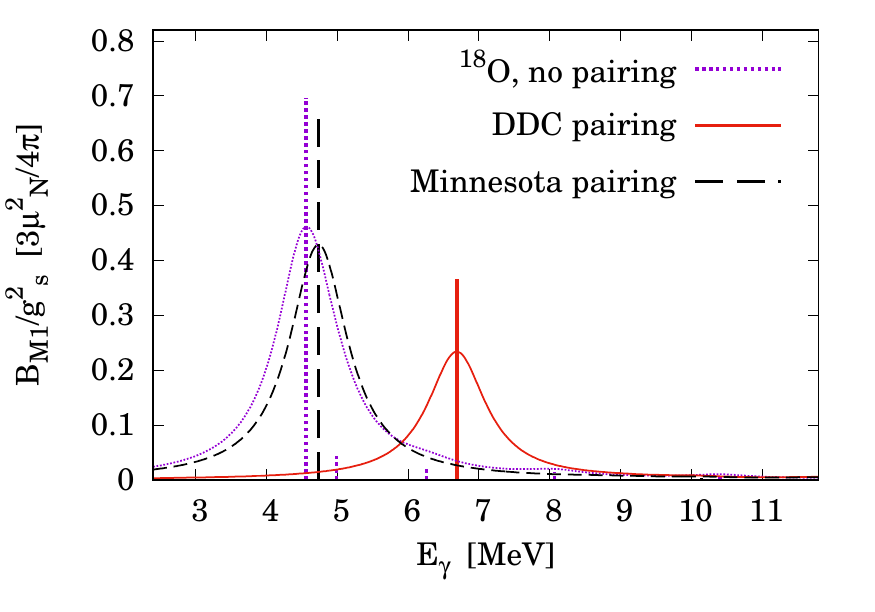}
  \caption{Discrete M1 transition strength for $^{18}$O. 
Note that $E_{\gamma}=E_f -E_{\rm GS}$. 
The continuous distribution is also plotted 
by smearing the discrete strength with a Cauchy-Lorenz profile, whose 
full width at half maximum (FWHM) is $1.0$ MeV.} \label{fig:1218_2018}
\end{center} \end{figure}

In Fig. \ref{fig:1218_2018}, we plot the M1 transition strength: 
\beq
 B_{\rm M1}(E_{\gamma}) = \sum_{\nu=0,\pm 1} \abs{\Braket{f (1^+) | \oprt{Q}_{\nu} | \Psi_{\rm GS} (0^+)}}^2, 
\eeq
where $E_{\gamma}=E_{f}-E_{\rm GS}$. 
Note that, because the system is spherical, $\nu=0$ and $\pm 1$ yield 
the same result. 
The figure also shows the continuous distribution obtained by smearing the 
discrete strength with a Cauchy-Lorenz profile, whose 
full width at half maximum (FWHM) is $1.0$ MeV. 

In Fig. \ref{fig:1218_2018}, the highest transition strength without pairing locates at 
$E_{\gamma} \cong 4.6$ MeV, which coincides well with the gap energy between 
the SP states $1d_{5/2}$ and $1d_{3/2}$.
Thus, the transition $1d_{5/2} \longrightarrow 1d_{3/2}$ 
plays a major role in this case.

The numerical SRV can be obtained from the sum of discrete M1 transition strength,
resulting in,
\beq
 S_{\rm M1,cal.} = \sum_{E_{\gamma}} B_{\rm M1}(E_{\gamma}) \cong 0.799 g^2_s.
\eeq
On the other side, the corresponding analytic SRV for the $S_{12}=1$ component reads from Eq. (\ref{eq:SRV_ANA}),
\beq
 S_{\rm M1,NP}=2g^2_s N^{(d_{5/2},d_{5/2};J=0)}_{S_{12}=1}, \nonumber
\eeq
where the spin-triplet component $N^{(d_{5/2},d_{5/2};J=0)}_{S_{12}=1} =2/5$ exactly \cite{60Edm}, 
i.e. the analytic SRV equals $4/5g^2_s = 0.8g^2_s $. 
Thus, our numerical SRV is consistent to the analytic one. 
Notice also that, in Fig. \ref{fig:1218_2018}, the highest 
strength has $0.69g^2_s$, which exhausts about $87$\% of the total SRV. 
This again means a major component comes from $1d_{5/2} \longrightarrow 1d_{3/2}$ transition.

\subsubsection{pairing case}
In the following, the non-diagonal terms of Hamiltonian, 
$v_{NN}$ and $x_{\rm rec}$, are also taken into account in the analysis of $^{18}$O. 
For the pairing interaction, we employ a density-dependent contact (DDC) potential, 
similarly as in Refs. \cite{1991BE,1997EBH,2005HS,07Hagi_01,2010Oishi}. 
That is, 
\beqa
 v_{NN} (\bir_1 , \bir_2) &=& w(\abs{\bi{R}_{12}}) \cdot \delta(\bir_1-\bir_2), \nonumber \\
 w(r) &=& w_0 \left[ 1-f(r) \right],
\eeqa
where $\bi{R}_{12}=(\bir_1+\bir_2)/2$. 
The $f(r)$ is the same WS profile given in Eq. (\ref{eq:WS}), and 
it schematically describes the density-dependence of the 
effective pairing interaction. 
Its bare strength is determined consistently from the neutron-neutron 
scattering length in vacuum, $a_v=-18.5$ fm, 
within the cutoff energy \cite{1991BE,1997EBH}. 
That is, 
\beq
 w_0 = \frac{4\pi^2 \hbar^2 a_v}{m_{n}(\pi -2a_v k_{\rm cut})}~~~[{\rm MeV \cdot fm^3}],
\eeq
where $k_{\rm cut}=\sqrt{m_n E_{\rm cut}}/ \hbar$. 

The two-neutron GS is solved by diagonalizing the three-body 
Hamiltonian via the anti-symmetrized TP basis coupled to $0^+$. 
As the result, the GS energy for $^{18}$O is obtained, $E_{\rm GS}=-12.019$ MeV, 
which is in fair agreement with the experimental value, 
$-12.188$ MeV \cite{NNDCHP}. 
On the other side, the $1^+$ excited states, which satisfy 
$H \ket{f(1^+)} = E_f \ket{f(1^+)}$, 
are solved by diagonalizing the same Hamiltonian, but via the $1^+$ TP basis.

\begin{table}[tb] \begin{center}
\caption{Properties of the ground and $1^+$-excited states of $^{18}$O obtained with the three-body-model calculation. 
Note that $E_{\rm GS}=-12.188$ MeV in the experimental data \cite{NNDCHP}, 
with respect to the two-neutron-separation threshold. 
The sum-rule values denoted by SRV are determined as $2g^2_s N_{S_{12}=1}$, and 
shown in comparison to the calculated values $S_{\rm M1,cal.}$, which are obtained by integrating the $B_{\rm M1}(E_{\gamma})$ distribution. } \label{table:18O_GS}
  \catcode`? = \active \def?{\phantom{0}} 
  \begingroup \renewcommand{\arraystretch}{1.2}
  \begin{tabular*}{\hsize} { @{\extracolsep{\fill}} lrrr } \hline \hline
                          &DDC (full)?    &Minnesota?         &No pair.?  \\ \hline
  ?$E_{\rm GS}$             &$-12.019$ MeV? &$-12.013$ MeV?     &$-8.286$ MeV?  \\
  ?$\Braket{v_{NN}}$       &$-4.337$ MeV?  &$-4.022$ MeV?      &$0$ MeV?    \\
  ?$\Braket{x_{\rm rec}}$   &$-0.177$ MeV?  &$-0.404$ MeV?      &$0$ MeV?    \\
  ?$N_{d_{5/2}\otimes d_{5/2}}$  &$89.9$\%?     &$94.9$\%?           &$100$\%?  \\
  ?$N_{S_{12}=1}$            &$19.7$\%?      &$34.9$\%?           &$40$\%?  \\
                          &(numerical)?  &(numerical)?         &(analytic)?  \\
  ?SRV            &$0.394 g^2_s$?      &$0.698 g^2_s$?           &$0.8 g^2_s$?  \\
  &&& \\
  ?$E^{(1)}_{f}$           &$-5.329$ MeV? &$-7.286$ MeV?        &$-3.729$ MeV?  \\
  ?$S_{\rm M1,cal.}$        &$0.393 g^2_s$? &$0.696 g^2_s$?        &$0.799 g^2_s$?  \\  \hline \hline
  \end{tabular*}
  \endgroup
  \catcode`? = 12 
\end{center} \end{table}

In Table \ref{table:18O_GS}, properties of the $0^+$ ground state and the 
lowest $1^+$ state are summarized. 
One can read that the pairing correlation leads to the 
deeper binding of two neutrons. 
Furthermore, the GS cannot be pure $(d_{5/2})^2$ state, but it
includes other components. 
As the result, the GS spin-triplet ($S_{12}=1$) component is remarkably suppressed from 
the no-pairing case, similarly to the dineutron and diproton 
correlations \cite{2005HS,2010Oishi}. 
Note that this component is evaluated by numerically 
integrating the initial-state density, $\abs{\Psi_{\rm GS}(\bir_1,\bir_2)}^2$, 
but after the spin-triplet projection. 
That is, 
\beq
 N_{S_{12}=1} = \int d\bir_1 \int d\bir_2 \abs{\hat{P}_{S_{12}=1} \Psi_{\rm GS}(\bir_1,\bir_2)}^2. \label{eq:vayle}
\eeq
Notice also that the total density is normalized as 
$\int d\bir_1 \int d\bir_2 \abs{\Psi_{\rm GS}(\bir_1,\bir_2)}^2 = N_{S_{12}=0} +N_{S_{12}=1}=1$. 

In Fig. \ref{fig:1218_2018}, the M1 transition strength with the DDC pairing interaction is presented for $^{18}$O. 
In comparison to the no-pairing case, a significant suppression of the transition strength is obtained.
In Table \ref{table:18O_GS}, the lowest $1^+$ energy, $E^{(1)}_{f}$, is also displayed. 
Its change between the DDC and no-pairing cases is smaller 
than the corresponding change in the GS energy. 
Namely, compared with the GS energy, 
the excitation energies of $1^+$ states are less sensitive on 
the DDC pairing interaction. 
Thus, the main M1 peak is shifted to the higher-energy 
region at $E_{\gamma}\cong 6.7$ MeV.

In order to check the pairing effect more systematically, 
we repeat the same calculation but changing the absolute strength of the 
DDC-pairing interaction: $v_{NN} \longrightarrow f\cdot v_{NN}$, where 
$f=0.25,~0.50$, and $0.75$ are employed. 
From the result shown in Fig. \ref{fig:1218_2018}, 
one can clearly see that 
the M1 transition strength decreases with the 
increase of the pairing interaction. 
This is consistent to that the DDC-paring attraction enhances the 
spin-singlet component, which is not active for M1 excitation (for more 
details see the discussion in the following subsection). 
Also, because the energy level of the ground (excited) state is 
sensitive (insensitive) to the DDC interaction, 
the transition energy $E_{\gamma}$ increases when the attraction strength is enhanced.

The numerical sum-rule value can be obtained by integrating the 
strength distribution in Fig. \ref{fig:1218_2018}. 
The result in the full-DDC case is 
\beq
 S_{\rm M1,cal.} = \sum_{E_{\gamma}} B_{\rm M1}(E_{\gamma}) \cong 0.393 g^2_s. 
\eeq 
This value is, as expected from Eq. (\ref{eq:SRV_2n}), 
consistent with the initial $S_{12}=1$ component, 
$S_{\rm M1}({\rm 2n})=2g^2_s N_{S_{12}=1}$, where 
$N_{S_{12}=1}=0.197$ as evaluated by Eq. (\ref{eq:vayle}). 
Corresponding to the reduced M1 transition strength, 
the SRV is also suppressed by the DDC pairing correlation. 

From the experimental side, there is no evidence of 
the low-lying $1^+$ state in $^{18}$O around $4-8$ MeV, as predicted in Fig. \ref{fig:1218_2018}. 
One possible reason is that our theoretical model does not 
quantitatively reproduce the $1^+$ excited states. 
Especially, the $1^+$ excitation energies may be sensitive to the 
pairing interaction model, which has, however, not been optimized 
specifically to this unnatural-parity transition case. 
One should mention that, from the experimental point of view, 
the M1 excitation of $^{18}$O may be rather minor 
and behind the present measurability, 
because the number of valence neutrons is only two. 
This is in contrast to some 
nuclei \cite{1970Moreh_M1_Exp,1985Lasz_M1_Exp,1988Lasz_M1_Exp,1990Rich,1993Kam,1984Bohle,1984Bohle_02,2003Fearick,2016Pai,2017Shizuma}, 
where $10$-$20$ valence nucleons can contribute to the M1 transition, 
and make its response sufficiently strong. 
For $^{18}$O with the minor M1 excitation, 
the selective detection of $1^+$ states is possibly demanding. 
One should consider, e.g. the dominance of the E1 mode, and 
the competition with the E2 mode. 
To extract the pure information on the M1 process from existing data, 
further efforts may be necessary. 
We leave these issues for the future study.

\subsubsection{pairing-model dependence}
In the following we explore in more details the pairing-model dependence of the M1 excitations. 
For this purpose, the DDC pairing interaction is 
replaced to the Minnesota interaction \cite{77Thom}, which has been 
utilized in the similar three-body-model calculations \cite{01Myo,04Suzu,07Hagi_03,14Myo_Rev}. 
That is, 
\beqa
  && U_{\rm MIN} (\bir_1, \bir_2)  =v_r\exp \left(\frac{-d^2}{2q^2} \right) \nonumber \\
  && +v_s\exp \left(\frac{-d^2}{2\kappa^2_s q^2} \right) \hat{P}_{S_{12}=0} +v_t \left(\frac{-d^2}{2\kappa^2_t q^2} \right) \hat{P}_{S_{12}=1},
\eeqa
where $d\equiv \abs{\bir_1 -\bir_2}$, $v_r =200$ MeV, 
$v_s =-91.85$ MeV, $v_t =-178$ MeV, $q=0.5799$ fm, 
$\kappa_s =1.788$, and $\kappa_t =1.525$, as given in the original paper \cite{77Thom}. 
Here we also assume $u=1$ for Eq. (9) in Ref. \cite{77Thom}. 
The first term indicates a repulsive core, whereas the second (third) term 
describes the attractive force in the spin-singlet (triplet) channel. 
For the present calculation of $^{18}$O, however, 
we need to use the 3\% enhancement, $f=1.03$, in order 
to reproduce the two-neutron separation energy: 
\beq
 v_{NN} (\bir_1, \bir_2) = f\cdot U_{\rm MIN} (\bir_1, \bir_2). 
\eeq
Note also that the core-neutron WS potential and 
the cutoff parameters are common to the previous case.

Before going to the result, we describe a comparison between the 
matrix elements of the DDC and Minnesota interactions. 
For our basis, anti-symmetrized TP states, $\left\{ \ket{\tilde{\Psi}_{k_1 k_2} (J^P)} \right\}$ 
coupled to the spin-parity $J^P=0^+$ and $1^+$, are employed. 
This can be decomposed into the spin-singlet and triplet parts: 
\beqa
 \ket{\tilde{\Psi}_{k_1 k_2} (J^P)} &=& \alpha^{(0)}_{k_1 k_2} \ket{\tilde{\Psi}_{k_1 k_2} (J^P,S_{12}=0)} \nonumber \\
 & & +\alpha^{(1)}_{k_1 k_2} \ket{\tilde{\Psi}_{k_1 k_2} (J^P,S_{12}=1)}. 
\eeqa
Thus, the matrix element of the pairing interaction can also be decomposed as 
\beqa
 && \Braket{\tilde{\Psi}_{k'_1 k'_2} (J^P) \mid v_{NN} \mid \tilde{\Psi}_{k_1 k_2} (J^P)} \nonumber \\
 && =\alpha^{(0)*}_{k'_1 k'_2} \alpha^{(0)}_{k_1 k_2} \Braket{v_{NN}}_{(J^P,S_{12}=0)~k'_1 k'_2 k_1 k_2} \nonumber \\
 && ~~+\alpha^{(1)*}_{k'_1 k'_2} \alpha^{(1)}_{k_1 k_2} \Braket{v_{NN}}_{(J^P,S_{12}=1)~k'_1 k'_2 k_1 k_2}. 
\eeqa
For the zero-range DDC pairing, $v_{NN} \propto \delta(\bir_2-\bir_1)$, only 
the $S_{12}=0$ ($S_{12}=1$) term survives for $J^P=0^+$ ($1^+$) \cite{2012Tani}. 
Therefore, the zero-range attraction enhances only 
the $S_{12}=0$ ($S_{12}=1$) component in the $J^P=0^+$ ($1^+$) state. 
On the other side, if $v_{NN}$ has a finite range, both terms 
may become non-zero for $J^P=0^+$ and $1^+$. 

In Table \ref{table:18O_GS}, our results with the Minnesota neutron-neutron interaction are summarized. 
First, one can find that the spin-triplet component, $N_{S_{12}=1}$, in the 
GS is slightly changed from the no-pairing case, but not as 
much as in the DDC-pairing case. 
In coincidence, the numerical SRV, $S_{\rm M1,cal.}$, 
has a larger value when compared with the DDC-pairing case. 
This difference between the two pairing models is understood by their ranges. 
Because the Minnesota model has a finite range, for the $0^+$ GS, 
it can contribute both in the $S_{12}=0$ and $S_{12}=1$ channels, 
whereas the zero-range DDC model only enhances the $S_{12}=0$ component.

Similarly to the GS result, the $1^+$ excitation of $^{18}$O 
depends on the choice of the pairing model: 
the lowest $1^+$ energy, $E^{(1)}_{f}$, shows a remarkable 
difference between the DDC and Minnesota cases. 
This is again a result of the finite range of the Minnesota interaction. 
With a finite range, its matrix element for the $1^+$ TP states 
can be larger than in the case of the zero-range DDC pairing. 

In Fig. \ref{fig:1218_2018}, the M1-transition 
strength by the Minnesota pairing is displayed, in comparison 
to the DDC pairing and no-pairing case. 
The result for the Minnesota pairing seems rather similar to that of the no-pairing case, 
in contrast to the DDC-pairing case. 
The transition energy, $E_{\gamma}=E_f-E_{\rm GS}$, is only slightly changed from the 
no-pairing result, because both the GS and excited energies are shifted by the Minnesota interaction. 
The $B_{\rm M1}$ value is slightly decreased, 
consistently to that the $S_{12}=1$ component in the GS is 
reduced by the Minnesota pairing. 

Consequently, although the pairing models are equivalently fitted 
to reproduce the standard GS energy, it does not guarantee the 
same prediction for the M1 excitation. 
It suggests that the M1 excitation data can be 
good reference observables to optimize the existing pairing models. 
Further developments to observe the minor, low-lying M1 excitation are now on 
demand from this point of view.

\subsection{mirror nucleus $^{18}$Ne}
Next we investigate the M1 excitations of the 
mirror nucleus, $^{18}$Ne, as the $^{16}$O core plus two valence 
protons with the Coulomb repulsion. 
Some parameters in the three-body model are revised to take 
the mass $m_p$, different $g$ factors, and Coulomb repulsion into account. 
For the Coulomb repulsion, 
the same procedure as in Refs. \cite{2010Oishi,2011Oishi} is utilized. 
That is, for the core-proton subsystem, we additionally employ the 
Coulomb potential of an uniformly-charged sphere. 
Also, for the proton-proton interaction, 
the factor $f=1.085$ is used for the DDC pairing 
in order to reproduce the empirical two-proton-separation energy. 
Namely, $v_{pp}=f\cdot v_{NN}+ e^2/\abs{\bir_2- \bir_1}$, combined with the Coulomb repulsion. 
Other parameters are kept unchanged from the $^{18}$O case.

\begin{table}[tb] \begin{center}
\caption{Same as Table \ref{table:18O_GS}, but for $^{18}$Ne. 
Note that $E_{\rm GS}=-4.523$ MeV in the experimental data \cite{NNDCHP}, 
with respect to the two-proton-separation threshold. } \label{table:18Ne_GS}
  \catcode`? = \active \def?{\phantom{0}} 
  \begingroup \renewcommand{\arraystretch}{1.2}
  \begin{tabular*}{\hsize} { @{\extracolsep{\fill}} lrrr } \hline \hline
                           &DDC?               &Minnesota?     &No pair.?  \\ \hline
  ?$E_{\rm GS}$              &$-4.527$ MeV?      &$-4.524$ MeV?  &$-1.154$ MeV?  \\
  ?$\Braket{v_{pp}}$        &$-4.015$ MeV?       &$-3.683$ MeV?  &$0$ MeV?    \\
  ?$\Braket{x_{\rm rec}}$    &$-0.127$ MeV?       &$-0.352$ MeV?  &$0$ MeV?    \\
  ?$N_{d_{5/2}\otimes d_{5/2}}$  &$88.5$\%?            &$93.5$\%?      &$100$\%?  \\
  ?$N_{S_{12}=1}$                &$18.9$\%?        &$33.0$\%?      &$40$\%?  \\
                           &(numerical)?         &(numerical)?  &(analytic)? \\
  ?SRV                    &$0.378 (g_l-g_s)^2$?  &$0.66(g_l-g_s)^2$?  &$0.8 (g_l-g_s)^2$?  \\
  &&&  \\
  ?$S_{\rm M1,cal.}$         &$0.376(g_l-g_s)^2$?  &$0.656(g_l-g_s)^2$?  &$0.799(g_l-g_s)^2$? \\   \hline \hline
  \end{tabular*}
  \endgroup
  \catcode`? = 12 
\end{center} \end{table}

\begin{figure}[tb] \begin{center}
  \includegraphics[width = 0.95\hsize]{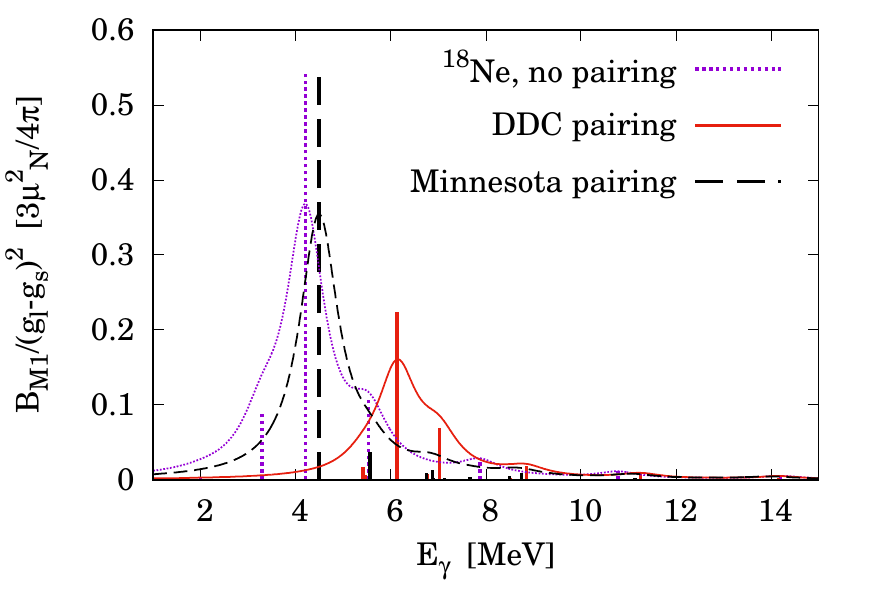}
  \caption{Same as Fig. \ref{fig:1218_2018}, but for $^{18}$Ne. Note that $E_{\gamma}=E_f -E_{\rm GS}$. } \label{fig:1219_2018}
\end{center} \end{figure}

Our results are summarized in Fig. \ref{fig:1219_2018} 
and Table \ref{table:18Ne_GS}. 
Indeed, one can observe a similar behavior of the M1 transition strength, i.e., 
it is suppressed when the DDC pairing correlation exists. 
This conclusion coincides with the reduction of the spin-triplet component 
in the GS by the zero-range DDC pairing, as shown in Table \ref{table:18Ne_GS}. 


Next we move to the Minnesota-pairing case: 
$v_{pp}=f\cdot U_{\rm MIN}+ e^2/\abs{\bir_2- \bir_1}$, where $f=1.148$ in order 
to reproduce the two-proton-separation energy. 
Its results are also displayed in Fig. \ref{fig:1219_2018} 
and Table \ref{table:18Ne_GS}. 
From the comparison, one can observe the similar tendency as in the $^{18}$O case. 
Namely, the reduction of the $S_{12}=1$ component 
in the GS is smaller than that by the DDC pairing. 
In coincidence, the $B_{\rm M1}$ distribution shows only slightly different 
shape than the no-pairing result.

The numerical SRVs in the DDC, Minnesota, and no-pairing cases 
are obtained as 
$S_{\rm M1,cal.}=0.402(g_l-g_s)^2$, 
$0.656(g_l-g_s)^2$, and 
$0.799(g_l-g_s)^2$, 
respectively. 
Thus, the typical reduction factor between the 
pairing and no-pairing SRVs 
is similar in the isobaric analogue systems, $^{18}$O and $^{18}$Ne. 
We also note that the no-pairing SRV is consistent to the 
analytic solution, $4(g_l-g_s)^2/5$.

Comparing with the $^{18}$O result, $^{18}$Ne shows a 
widely fragmented shape of the M1 strength. 
This is because the $1^+$ state is not stable, but a 
resonance in the $^{18}$Ne case. 
The typical (resonance) energy $E_r(1^+)$ can be estimated from 
the mean transition energy $E_{\gamma}$ and the GS energy. 
From figures \ref{fig:1218_2018} \& \ref{fig:1219_2018} and 
tables \ref{table:18O_GS} \& \ref{table:18Ne_GS}, 
that is estimated, in e.g. the full-DDC pairing case, 
as $E_r(1^+) \cong +3.2$ ($-5.3$) MeV 
for $^{18}$Ne ($^{18}$O). 
This energy difference is naturally understood by 
the Coulomb repulsive force both in the core-proton and 
proton-proton subsystems. 
Consequently, as a product of the Coulomb repulsion, 
the M1-excited state of $^{18}$Ne may be unbound 
for the proton emission, and 
its M1 distribution becomes fragmented, originating from 
the non-zero width of the resonance state.

\subsection{$pf$-shell nucleus $^{42}$Ca}
In the last two cases, we investigated the M1 excitation 
of the nucleon pair from the $sd$ shell. 
In the following, we move toward the $pf$-shell system, 
namely, $^{42}$Ca with the $^{40}$Ca core. 
For the $^{40}$Ca nucleus, which is doubly shell-closed, 
there has been no $1^+$ state 
measured in the low-lying region \cite{NNDCHP}. 
Thus, $^{40}$Ca is suitable to our rigid-core assumption.

For numerical computation, we change some parameters as 
$A_C=40$, $V_0=-55.7$ MeV, and $U_{ls}=10.8$ MeVfm$^2$. 
Other parameters in $V_C$, $v_{NN}$, and cutoff parameters 
remain the same as in the $^{18}$O case. 
This setting fairly reproduces the single-neutron 
energies of $1f_{7/2}$ and $1f_{5/2}$ in $^{41}$Ca \cite{NNDCHP}. 
Due to the Pauli principle, we exclude the SP states up to 
$1d_{3/2}$, which are occupied by the core.

The GS properties of $^{42}$Ca are summarized in Table \ref{table:42Ca_GS}. 
Its GS energy obtained with the DDC or Minnesota-pairing 
interaction is in a fair agreement with the empirical value, 
$-19.843$ MeV \cite{NNDCHP}. 
In the Minnesota case, the enhancement factor $f=1.12$ 
is needed to reproduce the GS energy: $v_{NN}=f\cdot U_{\rm MIN}$. 

\begin{figure}[tb] \begin{center}
  \includegraphics[width = 0.95\hsize]{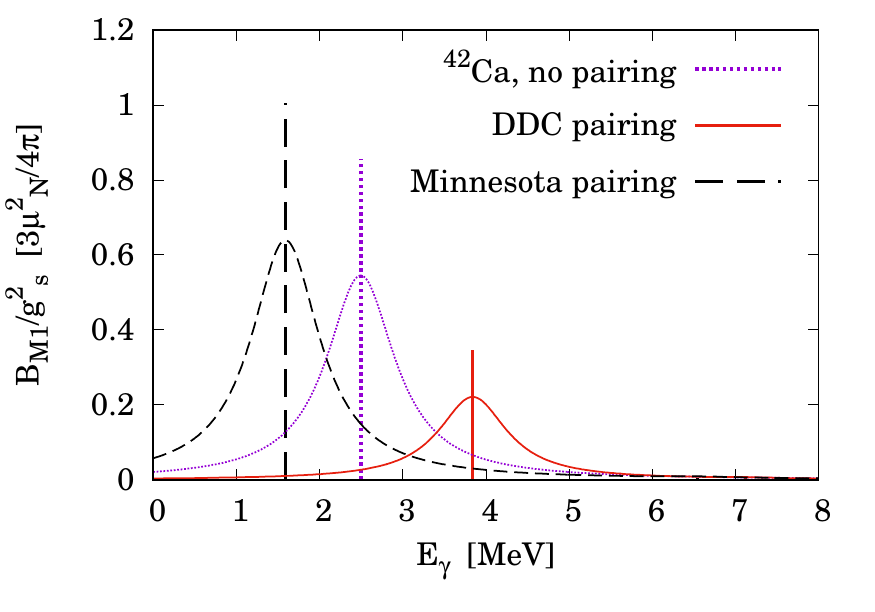}
  \caption{Same as Fig. \ref{fig:1218_2018}, but for $^{42}$Ca. 
Note that $E_{\gamma}=E_f -E_{\rm GS}$. } \label{fig:1221_2018}
\end{center} \end{figure}

In Fig. \ref{fig:1221_2018}, the M1 transition strength for 
$^{42}$Ca is shown for all the cases. 
First, comparing the DDC and no-pairing cases, 
one can find qualitatively the same conclusion as in the $sd$-shell nuclei: 
the DDC pairing suppresses the strength of M1 transitions. 
The SRV with the pairing is obtained as $S_{\rm M1,cal.}=0.352g^2_s$. 
This result is, as expected from Eq. (\ref{eq:SRV_2n}), 
consistent to the $S_{\rm M1}({\rm 2n})=2g^2_s N_{S_{12}=1}$, where 
$N_{S_{12}=1}=0.176$ as shown in Table \ref{table:42Ca_GS}. 
Similarly to the $sd$-shell case, 
the M1 SRV is shown to be linked with the 
coupled spin, which reflects the zero-range pairing effect.

In $^{42}$Ca, the Minnesota pairing provides 
a significant change from the other two cases. 
First the $S_{12}=1$ component in the GS is {\it enhanced} from 
the no-pairing result. 
This enhancement then leads to the increase of the 
$B_{\rm M1}$ and $S_{\rm M1,cal.}$ values, which can be 
consistent to our sum-rule formulation. 
The $1^+$-excitation energy is remarkably decreased, and thus, 
the transition energy, $E_{\gamma}$, locates 
at the lowest value among the three cases. 
These effects are understood from 
the finite range of the Minnesota force, similarly as 
explained in the $^{18}$O case. 
In this $^{42}$Ca case, however, the difference between 
the DDC and Minnesota models gets more significant, 
because of the $pf$-shell. 
This model dependence exists even when the DDC and Minnesota 
models are both fitted to the same GS energy.

\begin{table}[tb] \begin{center}
\caption{Same as Table \ref{table:18O_GS}, but for $^{42}$Ca. 
Note that $E_{\rm GS}=-19.843$ MeV in the experimental data \cite{NNDCHP}, 
with respect to the two-neutron-separation threshold. 
} \label{table:42Ca_GS}
  \catcode`? = \active \def?{\phantom{0}} 
  \begingroup \renewcommand{\arraystretch}{1.2}
  \begin{tabular*}{\hsize} { @{\extracolsep{\fill}} lrrr } \hline \hline
                         &DDC?           &Minnesota?       &No pair.?  \\ \hline
  ?$E_{\rm GS}$            &$-19.232$ MeV? &$-19.843$ MeV?   &$-16.795$ MeV?  \\
  ?$\Braket{v_{NN}}$      &$-2.999$ MeV?  &$-3.221$ MeV?    &$0$ MeV?    \\
  ?$\Braket{x_{\rm rec}}$  &$-0.005$ MeV?  &$-0.012$ MeV?    &$0$ MeV?    \\
  ?$N_{S_{12}=1}$           &$17.6$\%?      &$50.6$\%?        &$42.9$\%?  \\
                         &(numerical)?   &(numerical)?     &(analytic)? \\
  ?SRV                   &$0.352 g^2_s$?  &$1.012g^2_s$?    &$0.858 g^2_s$?  \\
  &&& \\
  ?$E^{(1)}_{f}$           &$-15.389$ MeV?  &$-18.253$ MeV?   &$-14.299$ MeV?  \\
  ?$S_{\rm M1,cal.}$        &$0.352 g^2_s$?  &$1.011 g^2_s$?    &$0.857 g^2_s$? \\   \hline \hline
  \end{tabular*}
  \endgroup
  \catcode`? = 12 
\end{center} \end{table}

\section{Summary} \label{Sec:Sum}
We have investigated M1 excitations in the systems with 
two-valence nucleons in the framework of the three-body model. 
First we have introduced model independent M1 sum rule, 
that is applicable to nuclei with two protons or neutrons above 
the core with the shell closure. 
We showed that the total sum of M1 transition strength can be linked directly to the spin-triplet component 
of the two valence nucleons in the shell. 
We also performed the three-body model calculations of M1 transition strength with 
the DDC and Minnesota-pairing interactions for $^{18}$O, $^{18}$Ne, and $^{42}$Ca. 
Model calculations accurately reproduced the proposed M1 sum rule values.

It is shown that the M1 excitation can be 
sensitive to the choice of pairing model. 
The $B_{\rm M1}$ and its SRV is enhanced or suppressed 
depending on the spin-triplet component in the GS, 
which depends on the pairing model. 
The same conclusion can apply both in the $sd$ and $pf$-shell nuclei. 
From these results, 
we expect that the M1 excitation is a promising 
probe to investigate the spin structure of valence nucleons, 
and/or to optimize the existing models for the pairing correlation. 

One should notice that the meson-exchange-current effect on the M1 transition 
has not been taken into account in this work. 
This effect can provide additional components of the 
M1-excitation strength \cite{1990Richter,2008Marcucci,1994Moraghe}, 
which may enhance the M1 SRV. 
For evaluation of this effect, we need to expand our model 
calculation to take the two-body and/or more-body components into account. 
To evaluate this effect, especially for the nuclides which 
have been discussed within the present three-body model, is one remaining task for future.

From the experimental side, additional studies are necessary to 
extract the spin information from the M1 excitation data, in particular for 
the systems with two-valence nucleons above the closed-shell core, 
where the introduced M1 sum rule could be validated. 
The main problem is the contribution from other, electric modes, 
which lead to the hindrance against the M1 strength. 
A close collaboration between theory and experiment may be 
necessary to resolve this issue.

\begin{acknowledgments}
T. Oishi sincerely thank T. O. Yamamoto and T. Uesaka for 
suggestions from the experimental side. 
This work is supported by the Croatian Science Foundation under the project Structure and Dynamics
of Exotic Femtosystems (IP-2014-09-9159) and by the QuantiXLie Centre of Excellence, 
a project co financed by the Croatian
Government and European Union through the European Regional Development Fund, the Competitiveness and Cohesion
Operational Programme (KK.01.1.1.01). 
\end{acknowledgments}



\begin{thebibliography}{74}%
\makeatletter
\providecommand \@ifxundefined [1]{%
 \@ifx{#1\undefined}
}%
\providecommand \@ifnum [1]{%
 \ifnum #1\expandafter \@firstoftwo
 \else \expandafter \@secondoftwo
 \fi
}%
\providecommand \@ifx [1]{%
 \ifx #1\expandafter \@firstoftwo
 \else \expandafter \@secondoftwo
 \fi
}%
\providecommand \natexlab [1]{#1}%
\providecommand \enquote  [1]{``#1''}%
\providecommand \bibnamefont  [1]{#1}%
\providecommand \bibfnamefont [1]{#1}%
\providecommand \citenamefont [1]{#1}%
\providecommand \href@noop [0]{\@secondoftwo}%
\providecommand \href [0]{\begingroup \@sanitize@url \@href}%
\providecommand \@href[1]{\@@startlink{#1}\@@href}%
\providecommand \@@href[1]{\endgroup#1\@@endlink}%
\providecommand \@sanitize@url [0]{\catcode `\\12\catcode `\$12\catcode
  `\&12\catcode `\#12\catcode `\^12\catcode `\_12\catcode `\%12\relax}%
\providecommand \@@startlink[1]{}%
\providecommand \@@endlink[0]{}%
\providecommand \url  [0]{\begingroup\@sanitize@url \@url }%
\providecommand \@url [1]{\endgroup\@href {#1}{\urlprefix }}%
\providecommand \urlprefix  [0]{URL }%
\providecommand \Eprint [0]{\href }%
\providecommand \doibase [0]{http://dx.doi.org/}%
\providecommand \selectlanguage [0]{\@gobble}%
\providecommand \bibinfo  [0]{\@secondoftwo}%
\providecommand \bibfield  [0]{\@secondoftwo}%
\providecommand \translation [1]{[#1]}%
\providecommand \BibitemOpen [0]{}%
\providecommand \bibitemStop [0]{}%
\providecommand \bibitemNoStop [0]{.\EOS\space}%
\providecommand \EOS [0]{\spacefactor3000\relax}%
\providecommand \BibitemShut  [1]{\csname bibitem#1\endcsname}%
\let\auto@bib@innerbib\@empty
\bibitem [{\citenamefont {Richter}(1985)}]{1985Rich}%
  \BibitemOpen
  \bibfield  {author} {\bibinfo {author} {\bibfnamefont {A.}~\bibnamefont
  {Richter}},\ }\href {\doibase https://doi.org/10.1016/0146-6410(85)90003-1}
  {\bibfield  {journal} {\bibinfo  {journal} {Progress in Particle and Nuclear
  Physics}\ }\textbf {\bibinfo {volume} {13}},\ \bibinfo {pages} {1 } (\bibinfo
  {year} {1985})}\BibitemShut {NoStop}%
\bibitem [{\citenamefont {Kneissl}\ \emph {et~al.}(1996)\citenamefont
  {Kneissl}, \citenamefont {Pitz},\ and\ \citenamefont
  {Zilges}}]{1996Kneis_Exp_Rev}%
  \BibitemOpen
  \bibfield  {author} {\bibinfo {author} {\bibfnamefont {U.}~\bibnamefont
  {Kneissl}}, \bibinfo {author} {\bibfnamefont {H.}~\bibnamefont {Pitz}}, \
  and\ \bibinfo {author} {\bibfnamefont {A.}~\bibnamefont {Zilges}},\ }\href
  {\doibase https://doi.org/10.1016/0146-6410(96)00055-5} {\bibfield  {journal}
  {\bibinfo  {journal} {Progress in Particle and Nuclear Physics}\ }\textbf
  {\bibinfo {volume} {37}},\ \bibinfo {pages} {349 } (\bibinfo {year}
  {1996})}\BibitemShut {NoStop}%
\bibitem [{\citenamefont {Pietralla}\ \emph {et~al.}(2008)\citenamefont
  {Pietralla}, \citenamefont {von Brentano},\ and\ \citenamefont
  {Lisetskiy}}]{2008Pietralla_Rev}%
  \BibitemOpen
  \bibfield  {author} {\bibinfo {author} {\bibfnamefont {N.}~\bibnamefont
  {Pietralla}}, \bibinfo {author} {\bibfnamefont {P.}~\bibnamefont {von
  Brentano}}, \ and\ \bibinfo {author} {\bibfnamefont {A.}~\bibnamefont
  {Lisetskiy}},\ }\href {\doibase https://doi.org/10.1016/j.ppnp.2007.08.002}
  {\bibfield  {journal} {\bibinfo  {journal} {Progress in Particle and Nuclear
  Physics}\ }\textbf {\bibinfo {volume} {60}},\ \bibinfo {pages} {225 }
  (\bibinfo {year} {2008})}\BibitemShut {NoStop}%
\bibitem [{\citenamefont {Heyde}\ \emph {et~al.}(2010)\citenamefont {Heyde},
  \citenamefont {von Neumann-Cosel},\ and\ \citenamefont
  {Richter}}]{2010Heyde_M1_Rev}%
  \BibitemOpen
  \bibfield  {author} {\bibinfo {author} {\bibfnamefont {K.}~\bibnamefont
  {Heyde}}, \bibinfo {author} {\bibfnamefont {P.}~\bibnamefont {von
  Neumann-Cosel}}, \ and\ \bibinfo {author} {\bibfnamefont {A.}~\bibnamefont
  {Richter}},\ }\href {\doibase 10.1103/RevModPhys.82.2365} {\bibfield
  {journal} {\bibinfo  {journal} {Rev. Mod. Phys.}\ }\textbf {\bibinfo {volume}
  {82}},\ \bibinfo {pages} {2365} (\bibinfo {year} {2010})},\ \bibinfo {note}
  {and references therein.}\BibitemShut {Stop}%
\bibitem [{\citenamefont {Bohle}\ \emph
  {et~al.}(1984{\natexlab{a}})\citenamefont {Bohle}, \citenamefont {Richter},
  \citenamefont {Steffen}, \citenamefont {Dieperink}, \citenamefont {Iudice},
  \citenamefont {Palumbo},\ and\ \citenamefont {Scholten}}]{1984Bohle}%
  \BibitemOpen
  \bibfield  {author} {\bibinfo {author} {\bibfnamefont {D.}~\bibnamefont
  {Bohle}}, \bibinfo {author} {\bibfnamefont {A.}~\bibnamefont {Richter}},
  \bibinfo {author} {\bibfnamefont {W.}~\bibnamefont {Steffen}}, \bibinfo
  {author} {\bibfnamefont {A.}~\bibnamefont {Dieperink}}, \bibinfo {author}
  {\bibfnamefont {N.~L.}\ \bibnamefont {Iudice}}, \bibinfo {author}
  {\bibfnamefont {F.}~\bibnamefont {Palumbo}}, \ and\ \bibinfo {author}
  {\bibfnamefont {O.}~\bibnamefont {Scholten}},\ }\href {\doibase
  https://doi.org/10.1016/0370-2693(84)91099-2} {\bibfield  {journal} {\bibinfo
   {journal} {Physics Letters B}\ }\textbf {\bibinfo {volume} {137}},\ \bibinfo
  {pages} {27 } (\bibinfo {year} {1984}{\natexlab{a}})}\BibitemShut {NoStop}%
\bibitem [{\citenamefont {Bohle}\ \emph
  {et~al.}(1984{\natexlab{b}})\citenamefont {Bohle}, \citenamefont {Küchler},
  \citenamefont {Richter},\ and\ \citenamefont {Steffen}}]{1984Bohle_02}%
  \BibitemOpen
  \bibfield  {author} {\bibinfo {author} {\bibfnamefont {D.}~\bibnamefont
  {Bohle}}, \bibinfo {author} {\bibfnamefont {G.}~\bibnamefont {Küchler}},
  \bibinfo {author} {\bibfnamefont {A.}~\bibnamefont {Richter}}, \ and\
  \bibinfo {author} {\bibfnamefont {W.}~\bibnamefont {Steffen}},\ }\href
  {\doibase https://doi.org/10.1016/0370-2693(84)90084-4} {\bibfield  {journal}
  {\bibinfo  {journal} {Physics Letters B}\ }\textbf {\bibinfo {volume}
  {148}},\ \bibinfo {pages} {260 } (\bibinfo {year}
  {1984}{\natexlab{b}})}\BibitemShut {NoStop}%
\bibitem [{\citenamefont {Richter}(1990)}]{1990Rich}%
  \BibitemOpen
  \bibfield  {author} {\bibinfo {author} {\bibfnamefont {A.}~\bibnamefont
  {Richter}},\ }\href {\doibase https://doi.org/10.1016/0375-9474(90)90571-3}
  {\bibfield  {journal} {\bibinfo  {journal} {Nuclear Physics A}\ }\textbf
  {\bibinfo {volume} {507}},\ \bibinfo {pages} {99 } (\bibinfo {year}
  {1990})}\BibitemShut {NoStop}%
\bibitem [{\citenamefont {Otsuka}(1990)}]{1990Taka}%
  \BibitemOpen
  \bibfield  {author} {\bibinfo {author} {\bibfnamefont {T.}~\bibnamefont
  {Otsuka}},\ }\href {\doibase https://doi.org/10.1016/0375-9474(90)90572-4}
  {\bibfield  {journal} {\bibinfo  {journal} {Nuclear Physics A}\ }\textbf
  {\bibinfo {volume} {507}},\ \bibinfo {pages} {129 } (\bibinfo {year}
  {1990})}\BibitemShut {NoStop}%
\bibitem [{\citenamefont {Kamerdzhiev}\ and\ \citenamefont
  {Speth}(1996)}]{1996Kamerdz}%
  \BibitemOpen
  \bibfield  {author} {\bibinfo {author} {\bibfnamefont {S.}~\bibnamefont
  {Kamerdzhiev}}\ and\ \bibinfo {author} {\bibfnamefont {J.}~\bibnamefont
  {Speth}},\ }\href {\doibase https://doi.org/10.1016/0375-9474(96)00080-2}
  {\bibfield  {journal} {\bibinfo  {journal} {Nuclear Physics A}\ }\textbf
  {\bibinfo {volume} {599}},\ \bibinfo {pages} {373 } (\bibinfo {year}
  {1996})},\ \bibinfo {note} {proceedings of the Groningnen Conference on Giant
  Resonances}\BibitemShut {NoStop}%
\bibitem [{\citenamefont {Vesely}\ \emph {et~al.}(2009)\citenamefont {Vesely},
  \citenamefont {Kvasil}, \citenamefont {Nesterenko}, \citenamefont {Kleinig},
  \citenamefont {Reinhard},\ and\ \citenamefont {Ponomarev}}]{2009Vesely}%
  \BibitemOpen
  \bibfield  {author} {\bibinfo {author} {\bibfnamefont {P.}~\bibnamefont
  {Vesely}}, \bibinfo {author} {\bibfnamefont {J.}~\bibnamefont {Kvasil}},
  \bibinfo {author} {\bibfnamefont {V.~O.}\ \bibnamefont {Nesterenko}},
  \bibinfo {author} {\bibfnamefont {W.}~\bibnamefont {Kleinig}}, \bibinfo
  {author} {\bibfnamefont {P.~G.}\ \bibnamefont {Reinhard}}, \ and\ \bibinfo
  {author} {\bibfnamefont {V.~Y.}\ \bibnamefont {Ponomarev}},\ }\href {\doibase
  10.1103/PhysRevC.80.031302(R)} {\bibfield  {journal} {\bibinfo  {journal} {Phys.
  Rev. C}\ }\textbf {\bibinfo {volume} {80}},\ \bibinfo {pages} {031302(R)}
  (\bibinfo {year} {2009})}\BibitemShut {NoStop}%
\bibitem [{\citenamefont {Nesterenko}\ \emph
  {et~al.}(2010{\natexlab{a}})\citenamefont {Nesterenko}, \citenamefont
  {Kvasil}, \citenamefont {Vesely}, \citenamefont {Kleinig}, \citenamefont
  {Reinhard},\ and\ \citenamefont {Ponomarev}}]{2010Nest}%
  \BibitemOpen
  \bibfield  {author} {\bibinfo {author} {\bibfnamefont {V.~O.}\ \bibnamefont
  {Nesterenko}}, \bibinfo {author} {\bibfnamefont {J.}~\bibnamefont {Kvasil}},
  \bibinfo {author} {\bibfnamefont {P.}~\bibnamefont {Vesely}}, \bibinfo
  {author} {\bibfnamefont {W.}~\bibnamefont {Kleinig}}, \bibinfo {author}
  {\bibfnamefont {P.-G.}\ \bibnamefont {Reinhard}}, \ and\ \bibinfo {author}
  {\bibfnamefont {V.~Y.}\ \bibnamefont {Ponomarev}},\ }\href
  {http://stacks.iop.org/0954-3899/37/i=6/a=064034} {\bibfield  {journal}
  {\bibinfo  {journal} {Journal of Physics G: Nuclear and Particle Physics}\
  }\textbf {\bibinfo {volume} {37}},\ \bibinfo {pages} {064034} (\bibinfo
  {year} {2010}{\natexlab{a}})}\BibitemShut {NoStop}%
\bibitem [{\citenamefont {Nesterenko}\ \emph
  {et~al.}(2010{\natexlab{b}})\citenamefont {Nesterenko}, \citenamefont
  {Kvasil}, \citenamefont {Vesely}, \citenamefont {Kleinig},\ and\
  \citenamefont {Reinhard}}]{2010Nest_2}%
  \BibitemOpen
  \bibfield  {author} {\bibinfo {author} {\bibfnamefont {V.~O.}\ \bibnamefont
  {Nesterenko}}, \bibinfo {author} {\bibfnamefont {J.}~\bibnamefont {Kvasil}},
  \bibinfo {author} {\bibfnamefont {P.}~\bibnamefont {Vesely}}, \bibinfo
  {author} {\bibfnamefont {W.}~\bibnamefont {Kleinig}}, \ and\ \bibinfo
  {author} {\bibfnamefont {P.-G.}\ \bibnamefont {Reinhard}},\ }\href {\doibase
  10.1142/S0218301310014972} {\bibfield  {journal} {\bibinfo  {journal} {Int.
  J. Mod. Phys. E}\ }\textbf {\bibinfo {volume} {19}},\ \bibinfo {pages} {558}
  (\bibinfo {year} {2010}{\natexlab{b}})}\BibitemShut {NoStop}%
\bibitem [{\citenamefont {Schwengner}\ \emph {et~al.}(2017)\citenamefont
  {Schwengner}, \citenamefont {Frauendorf},\ and\ \citenamefont
  {Brown}}]{2017Schwegner_M1}%
  \BibitemOpen
  \bibfield  {author} {\bibinfo {author} {\bibfnamefont {R.}~\bibnamefont
  {Schwengner}}, \bibinfo {author} {\bibfnamefont {S.}~\bibnamefont
  {Frauendorf}}, \ and\ \bibinfo {author} {\bibfnamefont {B.~A.}\ \bibnamefont
  {Brown}},\ }\href {\doibase 10.1103/PhysRevLett.118.092502} {\bibfield
  {journal} {\bibinfo  {journal} {Phys. Rev. Lett.}\ }\textbf {\bibinfo
  {volume} {118}},\ \bibinfo {pages} {092502} (\bibinfo {year}
  {2017})}\BibitemShut {NoStop}%
\bibitem [{\citenamefont {Balbutsev}\ \emph {et~al.}(2018)\citenamefont
  {Balbutsev}, \citenamefont {Molodtsova},\ and\ \citenamefont
  {Schuck}}]{2018Balb}%
  \BibitemOpen
  \bibfield  {author} {\bibinfo {author} {\bibfnamefont {E.~B.}\ \bibnamefont
  {Balbutsev}}, \bibinfo {author} {\bibfnamefont {I.~V.}\ \bibnamefont
  {Molodtsova}}, \ and\ \bibinfo {author} {\bibfnamefont {P.}~\bibnamefont
  {Schuck}},\ }\href {\doibase 10.1103/PhysRevC.97.044316} {\bibfield
  {journal} {\bibinfo  {journal} {Phys. Rev. C}\ }\textbf {\bibinfo {volume}
  {97}},\ \bibinfo {pages} {044316} (\bibinfo {year} {2018})}\BibitemShut
  {NoStop}%
\bibitem [{\citenamefont {Langanke}\ \emph {et~al.}(2004)\citenamefont
  {Langanke}, \citenamefont {Mart\'{\i}nez-Pinedo}, \citenamefont {von
  Neumann-Cosel},\ and\ \citenamefont {Richter}}]{2004Lang}%
  \BibitemOpen
  \bibfield  {author} {\bibinfo {author} {\bibfnamefont {K.}~\bibnamefont
  {Langanke}}, \bibinfo {author} {\bibfnamefont {G.}~\bibnamefont
  {Mart\'{\i}nez-Pinedo}}, \bibinfo {author} {\bibfnamefont {P.}~\bibnamefont
  {von Neumann-Cosel}}, \ and\ \bibinfo {author} {\bibfnamefont
  {A.}~\bibnamefont {Richter}},\ }\href {\doibase
  10.1103/PhysRevLett.93.202501} {\bibfield  {journal} {\bibinfo  {journal}
  {Phys. Rev. Lett.}\ }\textbf {\bibinfo {volume} {93}},\ \bibinfo {pages}
  {202501} (\bibinfo {year} {2004})}\BibitemShut {NoStop}%
\bibitem [{\citenamefont {Stone}\ and\ \citenamefont
  {Reinhard}(2007)}]{2007Stone}%
  \BibitemOpen
  \bibfield  {author} {\bibinfo {author} {\bibfnamefont {J.}~\bibnamefont
  {Stone}}\ and\ \bibinfo {author} {\bibfnamefont {P.-G.}\ \bibnamefont
  {Reinhard}},\ }\href {\doibase https://doi.org/10.1016/j.ppnp.2006.07.001}
  {\bibfield  {journal} {\bibinfo  {journal} {Progress in Particle and Nuclear
  Physics}\ }\textbf {\bibinfo {volume} {58}},\ \bibinfo {pages} {587 }
  (\bibinfo {year} {2007})}\BibitemShut {NoStop}%
\bibitem [{\citenamefont {Moreh}\ \emph {et~al.}(1970)\citenamefont {Moreh},
  \citenamefont {Shlomo},\ and\ \citenamefont {Wolf}}]{1970Moreh_M1_Exp}%
  \BibitemOpen
  \bibfield  {author} {\bibinfo {author} {\bibfnamefont {R.}~\bibnamefont
  {Moreh}}, \bibinfo {author} {\bibfnamefont {S.}~\bibnamefont {Shlomo}}, \
  and\ \bibinfo {author} {\bibfnamefont {A.}~\bibnamefont {Wolf}},\ }\href
  {\doibase 10.1103/PhysRevC.2.1144} {\bibfield  {journal} {\bibinfo  {journal}
  {Phys. Rev. C}\ }\textbf {\bibinfo {volume} {2}},\ \bibinfo {pages} {1144}
  (\bibinfo {year} {1970})}\BibitemShut {NoStop}%
\bibitem [{\citenamefont {Laszewski}\ \emph {et~al.}(1985)\citenamefont
  {Laszewski}, \citenamefont {Rullhusen}, \citenamefont {Hoblit},\ and\
  \citenamefont {LeBrun}}]{1985Lasz_M1_Exp}%
  \BibitemOpen
  \bibfield  {author} {\bibinfo {author} {\bibfnamefont {R.~M.}\ \bibnamefont
  {Laszewski}}, \bibinfo {author} {\bibfnamefont {P.}~\bibnamefont
  {Rullhusen}}, \bibinfo {author} {\bibfnamefont {S.~D.}\ \bibnamefont
  {Hoblit}}, \ and\ \bibinfo {author} {\bibfnamefont {S.~F.}\ \bibnamefont
  {LeBrun}},\ }\href {\doibase 10.1103/PhysRevLett.54.530} {\bibfield
  {journal} {\bibinfo  {journal} {Phys. Rev. Lett.}\ }\textbf {\bibinfo
  {volume} {54}},\ \bibinfo {pages} {530} (\bibinfo {year} {1985})}\BibitemShut
  {NoStop}%
\bibitem [{\citenamefont {Laszewski}\ \emph {et~al.}(1988)\citenamefont
  {Laszewski}, \citenamefont {Alarcon}, \citenamefont {Dale},\ and\
  \citenamefont {Hoblit}}]{1988Lasz_M1_Exp}%
  \BibitemOpen
  \bibfield  {author} {\bibinfo {author} {\bibfnamefont {R.~M.}\ \bibnamefont
  {Laszewski}}, \bibinfo {author} {\bibfnamefont {R.}~\bibnamefont {Alarcon}},
  \bibinfo {author} {\bibfnamefont {D.~S.}\ \bibnamefont {Dale}}, \ and\
  \bibinfo {author} {\bibfnamefont {S.~D.}\ \bibnamefont {Hoblit}},\ }\href
  {\doibase 10.1103/PhysRevLett.61.1710} {\bibfield  {journal} {\bibinfo
  {journal} {Phys. Rev. Lett.}\ }\textbf {\bibinfo {volume} {61}},\ \bibinfo
  {pages} {1710} (\bibinfo {year} {1988})}\BibitemShut {NoStop}%
\bibitem [{\citenamefont {Kamerdzhiev}\ \emph {et~al.}(1993)\citenamefont
  {Kamerdzhiev}, \citenamefont {Speth}, \citenamefont {Tertychny},\ and\
  \citenamefont {Wambach}}]{1993Kam}%
  \BibitemOpen
  \bibfield  {author} {\bibinfo {author} {\bibfnamefont {S.}~\bibnamefont
  {Kamerdzhiev}}, \bibinfo {author} {\bibfnamefont {J.}~\bibnamefont {Speth}},
  \bibinfo {author} {\bibfnamefont {G.}~\bibnamefont {Tertychny}}, \ and\
  \bibinfo {author} {\bibfnamefont {J.}~\bibnamefont {Wambach}},\ }\href@noop
  {} {\bibfield  {journal} {\bibinfo  {journal} {Zeitschrift f\"{u}r Physik A}\
  }\textbf {\bibinfo {volume} {346}},\ \bibinfo {pages} {253} (\bibinfo {year}
  {1993})}\BibitemShut {NoStop}%
\bibitem [{\citenamefont {Fearick}\ \emph {et~al.}(2003)\citenamefont
  {Fearick}, \citenamefont {Hartung}, \citenamefont {Langanke}, \citenamefont
  {Martínez-Pinedo}, \citenamefont {von Neumann-Cosel},\ and\ \citenamefont
  {Richter}}]{2003Fearick}%
  \BibitemOpen
  \bibfield  {author} {\bibinfo {author} {\bibfnamefont {R.}~\bibnamefont
  {Fearick}}, \bibinfo {author} {\bibfnamefont {G.}~\bibnamefont {Hartung}},
  \bibinfo {author} {\bibfnamefont {K.}~\bibnamefont {Langanke}}, \bibinfo
  {author} {\bibfnamefont {G.}~\bibnamefont {Martínez-Pinedo}}, \bibinfo
  {author} {\bibfnamefont {P.}~\bibnamefont {von Neumann-Cosel}}, \ and\
  \bibinfo {author} {\bibfnamefont {A.}~\bibnamefont {Richter}},\ }\href
  {\doibase https://doi.org/10.1016/j.nuclphysa.2003.07.016} {\bibfield
  {journal} {\bibinfo  {journal} {Nuclear Physics A}\ }\textbf {\bibinfo
  {volume} {727}},\ \bibinfo {pages} {41 } (\bibinfo {year}
  {2003})}\BibitemShut {NoStop}%
\bibitem [{\citenamefont {Pai}\ \emph {et~al.}(2016)\citenamefont {Pai},
  \citenamefont {Beck}, \citenamefont {Beller}, \citenamefont {Beyer},
  \citenamefont {Bhike}, \citenamefont {Derya}, \citenamefont {Gayer},
  \citenamefont {Isaak}, \citenamefont {Krishichayan}, \citenamefont {Kvasil},
  \citenamefont {L\"oher}, \citenamefont {Nesterenko}, \citenamefont
  {Pietralla}, \citenamefont {Mart\'{\i}nez-Pinedo}, \citenamefont {Mertes},
  \citenamefont {Ponomarev}, \citenamefont {Reinhard}, \citenamefont {Repko},
  \citenamefont {Ries}, \citenamefont {Romig}, \citenamefont {Savran},
  \citenamefont {Schwengner}, \citenamefont {Tornow}, \citenamefont {Werner},
  \citenamefont {Wilhelmy}, \citenamefont {Zilges},\ and\ \citenamefont
  {Zweidinger}}]{2016Pai}%
  \BibitemOpen
  \bibfield  {author} {\bibinfo {author} {\bibfnamefont {H.}~\bibnamefont
  {Pai}}, \bibinfo {author} {\bibfnamefont {T.}~\bibnamefont {Beck}}, \bibinfo
  {author} {\bibfnamefont {J.}~\bibnamefont {Beller}}, \bibinfo {author}
  {\bibfnamefont {R.}~\bibnamefont {Beyer}}, \bibinfo {author} {\bibfnamefont
  {M.}~\bibnamefont {Bhike}}, \bibinfo {author} {\bibfnamefont
  {V.}~\bibnamefont {Derya}}, \bibinfo {author} {\bibfnamefont
  {U.}~\bibnamefont {Gayer}}, \bibinfo {author} {\bibfnamefont
  {J.}~\bibnamefont {Isaak}}, \bibinfo {author} {\bibnamefont {Krishichayan}},
  \bibinfo {author} {\bibfnamefont {J.}~\bibnamefont {Kvasil}}, \bibinfo
  {author} {\bibfnamefont {B.}~\bibnamefont {L\"oher}}, \bibinfo {author}
  {\bibfnamefont {V.~O.}\ \bibnamefont {Nesterenko}}, \bibinfo {author}
  {\bibfnamefont {N.}~\bibnamefont {Pietralla}}, \bibinfo {author}
  {\bibfnamefont {G.}~\bibnamefont {Mart\'{\i}nez-Pinedo}}, \bibinfo {author}
  {\bibfnamefont {L.}~\bibnamefont {Mertes}}, \bibinfo {author} {\bibfnamefont
  {V.~Y.}\ \bibnamefont {Ponomarev}}, \bibinfo {author} {\bibfnamefont {P.-G.}\
  \bibnamefont {Reinhard}}, \bibinfo {author} {\bibfnamefont {A.}~\bibnamefont
  {Repko}}, \bibinfo {author} {\bibfnamefont {P.~C.}\ \bibnamefont {Ries}},
  \bibinfo {author} {\bibfnamefont {C.}~\bibnamefont {Romig}}, \bibinfo
  {author} {\bibfnamefont {D.}~\bibnamefont {Savran}}, \bibinfo {author}
  {\bibfnamefont {R.}~\bibnamefont {Schwengner}}, \bibinfo {author}
  {\bibfnamefont {W.}~\bibnamefont {Tornow}}, \bibinfo {author} {\bibfnamefont
  {V.}~\bibnamefont {Werner}}, \bibinfo {author} {\bibfnamefont
  {J.}~\bibnamefont {Wilhelmy}}, \bibinfo {author} {\bibfnamefont
  {A.}~\bibnamefont {Zilges}}, \ and\ \bibinfo {author} {\bibfnamefont
  {M.}~\bibnamefont {Zweidinger}},\ }\href {\doibase
  10.1103/PhysRevC.93.014318} {\bibfield  {journal} {\bibinfo  {journal} {Phys.
  Rev. C}\ }\textbf {\bibinfo {volume} {93}},\ \bibinfo {pages} {014318}
  (\bibinfo {year} {2016})}\BibitemShut {NoStop}%
\bibitem [{\citenamefont {Shizuma}\ \emph {et~al.}(2017)\citenamefont
  {Shizuma}, \citenamefont {Hayakawa}, \citenamefont {Daito}, \citenamefont
  {Ohgaki}, \citenamefont {Miyamoto},\ and\ \citenamefont
  {Minato}}]{2017Shizuma}%
  \BibitemOpen
  \bibfield  {author} {\bibinfo {author} {\bibfnamefont {T.}~\bibnamefont
  {Shizuma}}, \bibinfo {author} {\bibfnamefont {T.}~\bibnamefont {Hayakawa}},
  \bibinfo {author} {\bibfnamefont {I.}~\bibnamefont {Daito}}, \bibinfo
  {author} {\bibfnamefont {H.}~\bibnamefont {Ohgaki}}, \bibinfo {author}
  {\bibfnamefont {S.}~\bibnamefont {Miyamoto}}, \ and\ \bibinfo {author}
  {\bibfnamefont {F.}~\bibnamefont {Minato}},\ }\href {\doibase
  10.1103/PhysRevC.96.044316} {\bibfield  {journal} {\bibinfo  {journal} {Phys.
  Rev. C}\ }\textbf {\bibinfo {volume} {96}},\ \bibinfo {pages} {044316}
  (\bibinfo {year} {2017})}\BibitemShut {NoStop}%
\bibitem [{\citenamefont {Brink}\ and\ \citenamefont {Broglia}(2005)}]{05BB}%
  \BibitemOpen
  \bibfield  {author} {\bibinfo {author} {\bibfnamefont {D.}~\bibnamefont
  {Brink}}\ and\ \bibinfo {author} {\bibfnamefont {R.}~\bibnamefont
  {Broglia}},\ }\href {http://books.google.co.jp/books?id=K2HAuR-cQ4AC} {\emph
  {\bibinfo {title} {Nuclear Superfluidity: Pairing in Finite Systems}}},\
  Cambridge Monographs on Particle Physics, Nuclear Physics and Cosmology\
  (\bibinfo  {publisher} {Cambridge University Press},\ \bibinfo {address}
  {Cambridge, UK},\ \bibinfo {year} {2005})\BibitemShut {NoStop}%
\bibitem [{\citenamefont {Broglia}\ and\ \citenamefont
  {Zelevinsky}(2013)}]{13BZ}%
  \BibitemOpen
  \bibinfo {editor} {\bibfnamefont {R.~A.}\ \bibnamefont {Broglia}}\ and\
  \bibinfo {editor} {\bibfnamefont {V.}~\bibnamefont {Zelevinsky}},\ eds.,\
  \href {http://www.worldscientific.com/worldscibooks/10.1142/8526#t=aboutBook}
  {\emph {\bibinfo {title} {Fifty Years of Nuclear BCS: Pairing in Finite
  Systems}}}\ (\bibinfo  {publisher} {World Scientific, Singapore},\ \bibinfo
  {year} {2013})\BibitemShut {NoStop}%
\bibitem [{\citenamefont {Dean}\ and\ \citenamefont
  {Hjorth-Jensen}(2003)}]{03Dean_rev}%
  \BibitemOpen
  \bibfield  {author} {\bibinfo {author} {\bibfnamefont {D.~J.}\ \bibnamefont
  {Dean}}\ and\ \bibinfo {author} {\bibfnamefont {M.}~\bibnamefont
  {Hjorth-Jensen}},\ }\href {\doibase 10.1103/RevModPhys.75.607} {\bibfield
  {journal} {\bibinfo  {journal} {Rev. Mod. Phys.}\ }\textbf {\bibinfo {volume}
  {75}},\ \bibinfo {pages} {607} (\bibinfo {year} {2003})}\BibitemShut
  {NoStop}%
\bibitem [{\citenamefont {Bender}\ \emph {et~al.}(2003)\citenamefont {Bender},
  \citenamefont {Heenen},\ and\ \citenamefont {Reinhard}}]{03Bender_rev}%
  \BibitemOpen
  \bibfield  {author} {\bibinfo {author} {\bibfnamefont {M.}~\bibnamefont
  {Bender}}, \bibinfo {author} {\bibfnamefont {P.-H.}\ \bibnamefont {Heenen}},
  \ and\ \bibinfo {author} {\bibfnamefont {P.-G.}\ \bibnamefont {Reinhard}},\
  }\href {\doibase 10.1103/RevModPhys.75.121} {\bibfield  {journal} {\bibinfo
  {journal} {Rev. Mod. Phys.}\ }\textbf {\bibinfo {volume} {75}},\ \bibinfo
  {pages} {121} (\bibinfo {year} {2003})}\BibitemShut {NoStop}%
\bibitem [{\citenamefont {Hagino}\ and\ \citenamefont {Sagawa}(2005)}]{2005HS}%
  \BibitemOpen
  \bibfield  {author} {\bibinfo {author} {\bibfnamefont {K.}~\bibnamefont
  {Hagino}}\ and\ \bibinfo {author} {\bibfnamefont {H.}~\bibnamefont
  {Sagawa}},\ }\href {\doibase 10.1103/PhysRevC.72.044321} {\bibfield
  {journal} {\bibinfo  {journal} {Phys. Rev. C}\ }\textbf {\bibinfo {volume}
  {72}},\ \bibinfo {pages} {044321} (\bibinfo {year} {2005})}\BibitemShut
  {NoStop}%
\bibitem [{\citenamefont {Yoshida}(2009)}]{2009Yoshida_pyg}%
  \BibitemOpen
  \bibfield  {author} {\bibinfo {author} {\bibfnamefont {K.}~\bibnamefont
  {Yoshida}},\ }\href {\doibase 10.1103/PhysRevC.80.044324} {\bibfield
  {journal} {\bibinfo  {journal} {Phys. Rev. C}\ }\textbf {\bibinfo {volume}
  {80}},\ \bibinfo {pages} {044324} (\bibinfo {year} {2009})}\BibitemShut
  {NoStop}%
\bibitem [{\citenamefont {Terasaki}\ and\ \citenamefont
  {Engel}(2011)}]{2011TE}%
  \BibitemOpen
  \bibfield  {author} {\bibinfo {author} {\bibfnamefont {J.}~\bibnamefont
  {Terasaki}}\ and\ \bibinfo {author} {\bibfnamefont {J.}~\bibnamefont
  {Engel}},\ }\href {\doibase 10.1103/PhysRevC.84.014332} {\bibfield  {journal}
  {\bibinfo  {journal} {Phys. Rev. C}\ }\textbf {\bibinfo {volume} {84}},\
  \bibinfo {pages} {014332} (\bibinfo {year} {2011})}\BibitemShut {NoStop}%
\bibitem [{\citenamefont {Stetcu}\ \emph {et~al.}(2011)\citenamefont {Stetcu},
  \citenamefont {Bulgac}, \citenamefont {Magierski},\ and\ \citenamefont
  {Roche}}]{2011Stetcu}%
  \BibitemOpen
  \bibfield  {author} {\bibinfo {author} {\bibfnamefont {I.}~\bibnamefont
  {Stetcu}}, \bibinfo {author} {\bibfnamefont {A.}~\bibnamefont {Bulgac}},
  \bibinfo {author} {\bibfnamefont {P.}~\bibnamefont {Magierski}}, \ and\
  \bibinfo {author} {\bibfnamefont {K.~J.}\ \bibnamefont {Roche}},\ }\href
  {\doibase 10.1103/PhysRevC.84.051309} {\bibfield  {journal} {\bibinfo
  {journal} {Phys. Rev. C}\ }\textbf {\bibinfo {volume} {84}},\ \bibinfo
  {pages} {051309(R)} (\bibinfo {year} {2011})}\BibitemShut {NoStop}%
\bibitem [{\citenamefont {Ebata}\ \emph {et~al.}(2014)\citenamefont {Ebata},
  \citenamefont {Nakatsukasa},\ and\ \citenamefont {Inakura}}]{2014Ebata_pyg}%
  \BibitemOpen
  \bibfield  {author} {\bibinfo {author} {\bibfnamefont {S.}~\bibnamefont
  {Ebata}}, \bibinfo {author} {\bibfnamefont {T.}~\bibnamefont {Nakatsukasa}},
  \ and\ \bibinfo {author} {\bibfnamefont {T.}~\bibnamefont {Inakura}},\ }\href
  {\doibase 10.1103/PhysRevC.90.024303} {\bibfield  {journal} {\bibinfo
  {journal} {Phys. Rev. C}\ }\textbf {\bibinfo {volume} {90}},\ \bibinfo
  {pages} {024303} (\bibinfo {year} {2014})}\BibitemShut {NoStop}%
\bibitem [{\citenamefont {Dang}\ and\ \citenamefont {Hung}(2013)}]{2013Dang}%
  \BibitemOpen
  \bibfield  {author} {\bibinfo {author} {\bibfnamefont {N.~D.}\ \bibnamefont
  {Dang}}\ and\ \bibinfo {author} {\bibfnamefont {N.~Q.}\ \bibnamefont
  {Hung}},\ }\href {\doibase 10.1088/0954-3899/40/10/105103} {\bibfield
  {journal} {\bibinfo  {journal} {Journal of Physics G: Nuclear and Particle
  Physics}\ }\textbf {\bibinfo {volume} {40}},\ \bibinfo {pages} {105103}
  (\bibinfo {year} {2013})}\BibitemShut {NoStop}%
\bibitem [{\citenamefont {Oishi}\ \emph {et~al.}(2011)\citenamefont {Oishi},
  \citenamefont {Hagino},\ and\ \citenamefont {Sagawa}}]{2011Oishi}%
  \BibitemOpen
  \bibfield  {author} {\bibinfo {author} {\bibfnamefont {T.}~\bibnamefont
  {Oishi}}, \bibinfo {author} {\bibfnamefont {K.}~\bibnamefont {Hagino}}, \
  and\ \bibinfo {author} {\bibfnamefont {H.}~\bibnamefont {Sagawa}},\ }\href
  {\doibase 10.1103/PhysRevC.84.057301} {\bibfield  {journal} {\bibinfo
  {journal} {Phys. Rev. C}\ }\textbf {\bibinfo {volume} {84}},\ \bibinfo
  {pages} {057301} (\bibinfo {year} {2011})}\BibitemShut {NoStop}%
\bibitem [{\citenamefont {Eisenber}\ and\ \citenamefont
  {Greiner}(1970)}]{70Eisenberg}%
  \BibitemOpen
  \bibfield  {author} {\bibinfo {author} {\bibfnamefont {J.}~\bibnamefont
  {Eisenber}}\ and\ \bibinfo {author} {\bibfnamefont {W.}~\bibnamefont
  {Greiner}},\ }\href@noop {} {\emph {\bibinfo {title} {Nuclear Theory Volume
  2: Excitation Mechanisms of the Nucleus}}}\ (\bibinfo  {publisher}
  {North-Holland Publishing Company},\ \bibinfo {address} {Amsterdam},\
  \bibinfo {year} {1970})\BibitemShut {NoStop}%
\bibitem [{\citenamefont {Ring}\ and\ \citenamefont {Schuck}(1980)}]{80Ring}%
  \BibitemOpen
  \bibfield  {author} {\bibinfo {author} {\bibfnamefont {P.}~\bibnamefont
  {Ring}}\ and\ \bibinfo {author} {\bibfnamefont {P.}~\bibnamefont {Schuck}},\
  }\href@noop {} {\emph {\bibinfo {title} {The Nuclear Many-Body Problems}}}\
  (\bibinfo  {publisher} {Springer-Verlag},\ \bibinfo {address} {Berlin and
  Heidelberg, Germany},\ \bibinfo {year} {1980})\BibitemShut {NoStop}%
\bibitem [{\citenamefont {Wang}(1999)}]{1999Wang}%
  \BibitemOpen
  \bibfield  {author} {\bibinfo {author} {\bibfnamefont {S.}~\bibnamefont
  {Wang}},\ }\href {\doibase 10.1103/PhysRevA.60.262} {\bibfield  {journal}
  {\bibinfo  {journal} {Phys. Rev. A}\ }\textbf {\bibinfo {volume} {60}},\
  \bibinfo {pages} {262} (\bibinfo {year} {1999})}\BibitemShut {NoStop}%
\bibitem [{\citenamefont {Lu}\ and\ \citenamefont {Johnson}(2018)}]{2018Lu}%
  \BibitemOpen
  \bibfield  {author} {\bibinfo {author} {\bibfnamefont {Y.}~\bibnamefont
  {Lu}}\ and\ \bibinfo {author} {\bibfnamefont {C.~W.}\ \bibnamefont
  {Johnson}},\ }\href {\doibase 10.1103/PhysRevC.97.034330} {\bibfield
  {journal} {\bibinfo  {journal} {Phys. Rev. C}\ }\textbf {\bibinfo {volume}
  {97}},\ \bibinfo {pages} {034330} (\bibinfo {year} {2018})}\BibitemShut
  {NoStop}%
\bibitem [{\citenamefont {Ikeda}\ \emph {et~al.}(1963)\citenamefont {Ikeda},
  \citenamefont {Fujii},\ and\ \citenamefont {Fujita}}]{1963Ikeda}%
  \BibitemOpen
  \bibfield  {author} {\bibinfo {author} {\bibfnamefont {K.}~\bibnamefont
  {Ikeda}}, \bibinfo {author} {\bibfnamefont {S.}~\bibnamefont {Fujii}}, \ and\
  \bibinfo {author} {\bibfnamefont {J.~I.}\ \bibnamefont {Fujita}},\
  }\href@noop {} {\bibfield  {journal} {\bibinfo  {journal} {Phys. Lett.}\
  }\textbf {\bibinfo {volume} {3}},\ \bibinfo {pages} {271} (\bibinfo {year}
  {1963})}\BibitemShut {NoStop}%
\bibitem [{\citenamefont {Fujita}\ \emph {et~al.}(1964)\citenamefont {Fujita},
  \citenamefont {Fujii},\ and\ \citenamefont {Ikeda}}]{1964Fujita}%
  \BibitemOpen
  \bibfield  {author} {\bibinfo {author} {\bibfnamefont {J.~I.}\ \bibnamefont
  {Fujita}}, \bibinfo {author} {\bibfnamefont {S.}~\bibnamefont {Fujii}}, \
  and\ \bibinfo {author} {\bibfnamefont {K.}~\bibnamefont {Ikeda}},\
  }\href@noop {} {\bibfield  {journal} {\bibinfo  {journal} {Phys. Rev.}\
  }\textbf {\bibinfo {volume} {133}},\ \bibinfo {pages} {549} (\bibinfo {year}
  {1964})}\BibitemShut {NoStop}%
\bibitem [{\citenamefont {Johnson}(2015)}]{2015Johnson}%
  \BibitemOpen
  \bibfield  {author} {\bibinfo {author} {\bibfnamefont {C.~W.}\ \bibnamefont
  {Johnson}},\ }\href@noop {} {\bibfield  {journal} {\bibinfo  {journal} {Phys.
  Lett. B}\ }\textbf {\bibinfo {volume} {750}},\ \bibinfo {pages} {72}
  (\bibinfo {year} {2015})}\BibitemShut {NoStop}%
\bibitem [{\citenamefont {Hinohara}\ \emph {et~al.}(2015)\citenamefont
  {Hinohara}, \citenamefont {Kortelainen}, \citenamefont {Nazarewicz},\ and\
  \citenamefont {Olsen}}]{2015Hino}%
  \BibitemOpen
  \bibfield  {author} {\bibinfo {author} {\bibfnamefont {N.}~\bibnamefont
  {Hinohara}}, \bibinfo {author} {\bibfnamefont {M.}~\bibnamefont
  {Kortelainen}}, \bibinfo {author} {\bibfnamefont {W.}~\bibnamefont
  {Nazarewicz}}, \ and\ \bibinfo {author} {\bibfnamefont {E.}~\bibnamefont
  {Olsen}},\ }\href {\doibase 10.1103/PhysRevC.91.044323} {\bibfield  {journal}
  {\bibinfo  {journal} {Phys. Rev. C}\ }\textbf {\bibinfo {volume} {91}},\
  \bibinfo {pages} {044323} (\bibinfo {year} {2015})}\BibitemShut {NoStop}%
\bibitem [{\citenamefont {Bonasera}\ \emph {et~al.}(2018)\citenamefont
  {Bonasera}, \citenamefont {Anders},\ and\ \citenamefont
  {Shlomo}}]{2018Bonasera}%
  \BibitemOpen
  \bibfield  {author} {\bibinfo {author} {\bibfnamefont {G.}~\bibnamefont
  {Bonasera}}, \bibinfo {author} {\bibfnamefont {M.~R.}\ \bibnamefont
  {Anders}}, \ and\ \bibinfo {author} {\bibfnamefont {S.}~\bibnamefont
  {Shlomo}},\ }\href {\doibase 10.1103/PhysRevC.98.054316} {\bibfield
  {journal} {\bibinfo  {journal} {Phys. Rev. C}\ }\textbf {\bibinfo {volume}
  {98}},\ \bibinfo {pages} {054316} (\bibinfo {year} {2018})}\BibitemShut
  {NoStop}%
\bibitem [{\citenamefont {Garg}\ and\ \citenamefont {Colo}(2018)}]{2018Garg}%
  \BibitemOpen
  \bibfield  {author} {\bibinfo {author} {\bibfnamefont {U.}~\bibnamefont
  {Garg}}\ and\ \bibinfo {author} {\bibfnamefont {G.}~\bibnamefont {Colo}},\
  }\href@noop {} {\bibfield  {journal} {\bibinfo  {journal} {Prog. Part. Nucl.
  Phys}\ }\textbf {\bibinfo {volume} {101}},\ \bibinfo {pages} {55} (\bibinfo
  {year} {2018})}\BibitemShut {NoStop}%
\bibitem [{\citenamefont {Gupta}\ \emph {et~al.}(2018)\citenamefont {Gupta},
  \citenamefont {Howard}, \citenamefont {Garg}, \citenamefont {Matta},
  \citenamefont {\ifmmode \mbox{\c{S}}\else
  \c{S}\fi{}enyi\ifmmode~\breve{g}\else \u{g}\fi{}it}, \citenamefont {Itoh},
  \citenamefont {Ando}, \citenamefont {Aoki}, \citenamefont {Uchiyama},
  \citenamefont {Adachi}, \citenamefont {Fujiwara}, \citenamefont {Iwamoto},
  \citenamefont {Tamii}, \citenamefont {Akimune}, \citenamefont {Kadono},
  \citenamefont {Matsuda}, \citenamefont {Nakahara}, \citenamefont {Furuno},
  \citenamefont {Kawabata}, \citenamefont {Tsumura}, \citenamefont {Harakeh},\
  and\ \citenamefont {Kalantar-Nayestanaki}}]{2018Gupta}%
  \BibitemOpen
  \bibfield  {author} {\bibinfo {author} {\bibfnamefont {Y.~K.}\ \bibnamefont
  {Gupta}}, \bibinfo {author} {\bibfnamefont {K.~B.}\ \bibnamefont {Howard}},
  \bibinfo {author} {\bibfnamefont {U.}~\bibnamefont {Garg}}, \bibinfo {author}
  {\bibfnamefont {J.~T.}\ \bibnamefont {Matta}}, \bibinfo {author}
  {\bibfnamefont {M.}~\bibnamefont {\ifmmode \mbox{\c{S}}\else
  \c{S}\fi{}enyi\ifmmode~\breve{g}\else \u{g}\fi{}it}}, \bibinfo {author}
  {\bibfnamefont {M.}~\bibnamefont {Itoh}}, \bibinfo {author} {\bibfnamefont
  {S.}~\bibnamefont {Ando}}, \bibinfo {author} {\bibfnamefont {T.}~\bibnamefont
  {Aoki}}, \bibinfo {author} {\bibfnamefont {A.}~\bibnamefont {Uchiyama}},
  \bibinfo {author} {\bibfnamefont {S.}~\bibnamefont {Adachi}}, \bibinfo
  {author} {\bibfnamefont {M.}~\bibnamefont {Fujiwara}}, \bibinfo {author}
  {\bibfnamefont {C.}~\bibnamefont {Iwamoto}}, \bibinfo {author} {\bibfnamefont
  {A.}~\bibnamefont {Tamii}}, \bibinfo {author} {\bibfnamefont
  {H.}~\bibnamefont {Akimune}}, \bibinfo {author} {\bibfnamefont
  {C.}~\bibnamefont {Kadono}}, \bibinfo {author} {\bibfnamefont
  {Y.}~\bibnamefont {Matsuda}}, \bibinfo {author} {\bibfnamefont
  {T.}~\bibnamefont {Nakahara}}, \bibinfo {author} {\bibfnamefont
  {T.}~\bibnamefont {Furuno}}, \bibinfo {author} {\bibfnamefont
  {T.}~\bibnamefont {Kawabata}}, \bibinfo {author} {\bibfnamefont
  {M.}~\bibnamefont {Tsumura}}, \bibinfo {author} {\bibfnamefont {M.~N.}\
  \bibnamefont {Harakeh}}, \ and\ \bibinfo {author} {\bibfnamefont
  {N.}~\bibnamefont {Kalantar-Nayestanaki}},\ }\href {\doibase
  10.1103/PhysRevC.97.064323} {\bibfield  {journal} {\bibinfo  {journal} {Phys.
  Rev. C}\ }\textbf {\bibinfo {volume} {97}},\ \bibinfo {pages} {064323}
  (\bibinfo {year} {2018})}\BibitemShut {NoStop}%
\bibitem [{\citenamefont {Sagawa}\ and\ \citenamefont
  {Uesaka}(2016)}]{2016Sagawa}%
  \BibitemOpen
  \bibfield  {author} {\bibinfo {author} {\bibfnamefont {H.}~\bibnamefont
  {Sagawa}}\ and\ \bibinfo {author} {\bibfnamefont {T.}~\bibnamefont
  {Uesaka}},\ }\href {\doibase 10.1103/PhysRevC.94.064325} {\bibfield
  {journal} {\bibinfo  {journal} {Phys. Rev. C}\ }\textbf {\bibinfo {volume}
  {94}},\ \bibinfo {pages} {064325} (\bibinfo {year} {2016})}\BibitemShut
  {NoStop}%
\bibitem [{\citenamefont {Gambacurta}\ \emph {et~al.}(2016)\citenamefont
  {Gambacurta}, \citenamefont {Catara}, \citenamefont {Grasso}, \citenamefont
  {Sambataro}, \citenamefont {Andr\'{e}s},\ and\ \citenamefont
  {Lanza}}]{2016Gambacurta}%
  \BibitemOpen
  \bibfield  {author} {\bibinfo {author} {\bibfnamefont {D.}~\bibnamefont
  {Gambacurta}}, \bibinfo {author} {\bibfnamefont {F.}~\bibnamefont {Catara}},
  \bibinfo {author} {\bibfnamefont {M.}~\bibnamefont {Grasso}}, \bibinfo
  {author} {\bibfnamefont {M.}~\bibnamefont {Sambataro}}, \bibinfo {author}
  {\bibfnamefont {M.~V.}\ \bibnamefont {Andr\'{e}s}}, \ and\ \bibinfo {author}
  {\bibfnamefont {E.~G.}\ \bibnamefont {Lanza}},\ }\href {\doibase
  10.1103/PhysRevC.93.024309} {\bibfield  {journal} {\bibinfo  {journal} {Phys.
  Rev. C}\ }\textbf {\bibinfo {volume} {93}},\ \bibinfo {pages} {024309}
  (\bibinfo {year} {2016})}\BibitemShut {NoStop}%
\bibitem [{\citenamefont {Mason}\ \emph {et~al.}(2009)\citenamefont {Mason},
  \citenamefont {Chatterjee}, \citenamefont {Fortunato},\ and\ \citenamefont
  {Vitturi}}]{2009Mason}%
  \BibitemOpen
  \bibfield  {author} {\bibinfo {author} {\bibfnamefont {A.}~\bibnamefont
  {Mason}}, \bibinfo {author} {\bibfnamefont {R.}~\bibnamefont {Chatterjee}},
  \bibinfo {author} {\bibfnamefont {L.}~\bibnamefont {Fortunato}}, \ and\
  \bibinfo {author} {\bibfnamefont {A.}~\bibnamefont {Vitturi}},\ }\href
  {\doibase 10.1140/epja/i2008-10685-3} {\bibfield  {journal} {\bibinfo
  {journal} {The European Physical Journal A}\ }\textbf {\bibinfo {volume}
  {39}},\ \bibinfo {pages} {107} (\bibinfo {year} {2009})}\BibitemShut
  {NoStop}%
\bibitem [{\citenamefont {Osterfeld}(1992)}]{1992Osterfeld}%
  \BibitemOpen
  \bibfield  {author} {\bibinfo {author} {\bibfnamefont {F.}~\bibnamefont
  {Osterfeld}},\ }\href {\doibase 10.1103/RevModPhys.64.491} {\bibfield
  {journal} {\bibinfo  {journal} {Rev. Mod. Phys.}\ }\textbf {\bibinfo {volume}
  {64}},\ \bibinfo {pages} {491} (\bibinfo {year} {1992})}\BibitemShut
  {NoStop}%
\bibitem [{\citenamefont {Matsuo}\ \emph {et~al.}(2005)\citenamefont {Matsuo},
  \citenamefont {Mizuyama},\ and\ \citenamefont {Serizawa}}]{05Mats}%
  \BibitemOpen
  \bibfield  {author} {\bibinfo {author} {\bibfnamefont {M.}~\bibnamefont
  {Matsuo}}, \bibinfo {author} {\bibfnamefont {K.}~\bibnamefont {Mizuyama}}, \
  and\ \bibinfo {author} {\bibfnamefont {Y.}~\bibnamefont {Serizawa}},\ }\href
  {\doibase 10.1103/PhysRevC.71.064326} {\bibfield  {journal} {\bibinfo
  {journal} {Phys. Rev. C}\ }\textbf {\bibinfo {volume} {71}},\ \bibinfo
  {pages} {064326} (\bibinfo {year} {2005})}\BibitemShut {NoStop}%
\bibitem [{\citenamefont {Matsuo}\ \emph {et~al.}(2012)\citenamefont {Matsuo},
  \citenamefont {Shimoyama},\ and\ \citenamefont {Ootaki}}]{12Mats}%
  \BibitemOpen
  \bibfield  {author} {\bibinfo {author} {\bibfnamefont {M.}~\bibnamefont
  {Matsuo}}, \bibinfo {author} {\bibfnamefont {H.}~\bibnamefont {Shimoyama}}, \
  and\ \bibinfo {author} {\bibfnamefont {Y.}~\bibnamefont {Ootaki}},\ }\href
  {http://stacks.iop.org/1402-4896/2012/i=T150/a=014024} {\bibfield  {journal}
  {\bibinfo  {journal} {Physica Scripta}\ }\textbf {\bibinfo {volume} {T150}},\
  \bibinfo {pages} {014024} (\bibinfo {year} {2012})}\BibitemShut {NoStop}%
\bibitem [{\citenamefont {Bertulani}\ and\ \citenamefont
  {S.~Hussein}(2007)}]{07Bertulani_76}%
  \BibitemOpen
  \bibfield  {author} {\bibinfo {author} {\bibfnamefont {C.~A.}\ \bibnamefont
  {Bertulani}}\ and\ \bibinfo {author} {\bibfnamefont {M.}~\bibnamefont
  {S.~Hussein}},\ }\href {\doibase 10.1103/PhysRevC.76.051602} {\bibfield
  {journal} {\bibinfo  {journal} {Phys. Rev. C}\ }\textbf {\bibinfo {volume}
  {76}},\ \bibinfo {pages} {051602(R)} (\bibinfo {year} {2007})}\BibitemShut
  {NoStop}%
\bibitem [{\citenamefont {Hagino}\ \emph {et~al.}(2007)\citenamefont {Hagino},
  \citenamefont {Sagawa}, \citenamefont {Carbonell},\ and\ \citenamefont
  {Schuck}}]{07Hagi_01}%
  \BibitemOpen
  \bibfield  {author} {\bibinfo {author} {\bibfnamefont {K.}~\bibnamefont
  {Hagino}}, \bibinfo {author} {\bibfnamefont {H.}~\bibnamefont {Sagawa}},
  \bibinfo {author} {\bibfnamefont {J.}~\bibnamefont {Carbonell}}, \ and\
  \bibinfo {author} {\bibfnamefont {P.}~\bibnamefont {Schuck}},\ }\href
  {\doibase 10.1103/PhysRevLett.99.022506} {\bibfield  {journal} {\bibinfo
  {journal} {Phys. Rev. Lett.}\ }\textbf {\bibinfo {volume} {99}},\ \bibinfo
  {pages} {022506} (\bibinfo {year} {2007})}\BibitemShut {NoStop}%
\bibitem [{\citenamefont {Hagino}\ and\ \citenamefont
  {Sagawa}(2014)}]{14Hagi_2n}%
  \BibitemOpen
  \bibfield  {author} {\bibinfo {author} {\bibfnamefont {K.}~\bibnamefont
  {Hagino}}\ and\ \bibinfo {author} {\bibfnamefont {H.}~\bibnamefont
  {Sagawa}},\ }\href {\doibase 10.1103/PhysRevC.89.014331} {\bibfield
  {journal} {\bibinfo  {journal} {Phys. Rev. C}\ }\textbf {\bibinfo {volume}
  {89}},\ \bibinfo {pages} {014331} (\bibinfo {year} {2014})}\BibitemShut
  {NoStop}%
\bibitem [{\citenamefont {Oishi}\ \emph {et~al.}(2010)\citenamefont {Oishi},
  \citenamefont {Hagino},\ and\ \citenamefont {Sagawa}}]{2010Oishi}%
  \BibitemOpen
  \bibfield  {author} {\bibinfo {author} {\bibfnamefont {T.}~\bibnamefont
  {Oishi}}, \bibinfo {author} {\bibfnamefont {K.}~\bibnamefont {Hagino}}, \
  and\ \bibinfo {author} {\bibfnamefont {H.}~\bibnamefont {Sagawa}},\ }\href
  {\doibase 10.1103/PhysRevC.82.024315} {\bibfield  {journal} {\bibinfo
  {journal} {Phys. Rev. C}\ }\textbf {\bibinfo {volume} {82}},\ \bibinfo
  {pages} {024315} (\bibinfo {year} {2010})},\ \bibinfo {note} {with
  erratum.}\BibitemShut {Stop}%
\bibitem [{\citenamefont {Oishi}\ \emph {et~al.}(2014)\citenamefont {Oishi},
  \citenamefont {Hagino},\ and\ \citenamefont {Sagawa}}]{2014Oishi}%
  \BibitemOpen
  \bibfield  {author} {\bibinfo {author} {\bibfnamefont {T.}~\bibnamefont
  {Oishi}}, \bibinfo {author} {\bibfnamefont {K.}~\bibnamefont {Hagino}}, \
  and\ \bibinfo {author} {\bibfnamefont {H.}~\bibnamefont {Sagawa}},\ }\href
  {\doibase 10.1103/PhysRevC.90.034303} {\bibfield  {journal} {\bibinfo
  {journal} {Phys. Rev. C}\ }\textbf {\bibinfo {volume} {90}},\ \bibinfo
  {pages} {034303} (\bibinfo {year} {2014})}\BibitemShut {NoStop}%
\bibitem [{\citenamefont {Hagino}\ and\ \citenamefont
  {Sagawa}(2016)}]{2016Hagino}%
  \BibitemOpen
  \bibfield  {author} {\bibinfo {author} {\bibfnamefont {K.}~\bibnamefont
  {Hagino}}\ and\ \bibinfo {author} {\bibfnamefont {H.}~\bibnamefont
  {Sagawa}},\ }\href {\doibase 10.1007/s00601-015-1027-3} {\bibfield  {journal}
  {\bibinfo  {journal} {Few-Body Systems}\ }\textbf {\bibinfo {volume} {57}},\
  \bibinfo {pages} {185} (\bibinfo {year} {2016})}\BibitemShut {NoStop}%
\bibitem [{\citenamefont {Kikuchi}\ \emph {et~al.}(2016)\citenamefont
  {Kikuchi}, \citenamefont {Ogata}, \citenamefont {Kubota}, \citenamefont
  {Sasano},\ and\ \citenamefont {Uesaka}}]{2016Kiku}%
  \BibitemOpen
  \bibfield  {author} {\bibinfo {author} {\bibfnamefont {Y.}~\bibnamefont
  {Kikuchi}}, \bibinfo {author} {\bibfnamefont {K.}~\bibnamefont {Ogata}},
  \bibinfo {author} {\bibfnamefont {Y.}~\bibnamefont {Kubota}}, \bibinfo
  {author} {\bibfnamefont {M.}~\bibnamefont {Sasano}}, \ and\ \bibinfo {author}
  {\bibfnamefont {T.}~\bibnamefont {Uesaka}},\ }\href {\doibase
  10.1093/ptep/ptw148} {\bibfield  {journal} {\bibinfo  {journal} {Progress of
  Theoretical and Experimental Physics}\ }\textbf {\bibinfo {volume} {2016}},\
  \bibinfo {pages} {103D03} (\bibinfo {year} {2016})}\BibitemShut {NoStop}%
\bibitem [{\citenamefont {Grigorenko}\ \emph {et~al.}(2018)\citenamefont
  {Grigorenko}, \citenamefont {Vaagen},\ and\ \citenamefont
  {Zhukov}}]{2018Gri}%
  \BibitemOpen
  \bibfield  {author} {\bibinfo {author} {\bibfnamefont {L.~V.}\ \bibnamefont
  {Grigorenko}}, \bibinfo {author} {\bibfnamefont {J.~S.}\ \bibnamefont
  {Vaagen}}, \ and\ \bibinfo {author} {\bibfnamefont {M.~V.}\ \bibnamefont
  {Zhukov}},\ }\href {\doibase 10.1103/PhysRevC.97.034605} {\bibfield
  {journal} {\bibinfo  {journal} {Phys. Rev. C}\ }\textbf {\bibinfo {volume}
  {97}},\ \bibinfo {pages} {034605} (\bibinfo {year} {2018})},\ \bibinfo {note}
  {and references therein.}\BibitemShut {Stop}%
\bibitem [{\citenamefont {Suzuki}\ and\ \citenamefont
  {Ikeda}(1988)}]{88Suzuki_COSM}%
  \BibitemOpen
  \bibfield  {author} {\bibinfo {author} {\bibfnamefont {Y.}~\bibnamefont
  {Suzuki}}\ and\ \bibinfo {author} {\bibfnamefont {K.}~\bibnamefont {Ikeda}},\
  }\href {\doibase 10.1103/PhysRevC.38.410} {\bibfield  {journal} {\bibinfo
  {journal} {Phys. Rev. C}\ }\textbf {\bibinfo {volume} {38}},\ \bibinfo
  {pages} {410} (\bibinfo {year} {1988})}\BibitemShut {NoStop}%
\bibitem [{\citenamefont {Bertsch}\ and\ \citenamefont
  {Esbensen}(1991)}]{1991BE}%
  \BibitemOpen
  \bibfield  {author} {\bibinfo {author} {\bibfnamefont {G.}~\bibnamefont
  {Bertsch}}\ and\ \bibinfo {author} {\bibfnamefont {H.}~\bibnamefont
  {Esbensen}},\ }\href {\doibase
  http://dx.doi.org/10.1016/0003-4916(91)90033-5} {\bibfield  {journal}
  {\bibinfo  {journal} {Annals of Physics}\ }\textbf {\bibinfo {volume}
  {209}},\ \bibinfo {pages} {327 } (\bibinfo {year} {1991})}\BibitemShut
  {NoStop}%
\bibitem [{\citenamefont {Esbensen}\ \emph {et~al.}(1997)\citenamefont
  {Esbensen}, \citenamefont {Bertsch},\ and\ \citenamefont
  {Hencken}}]{1997EBH}%
  \BibitemOpen
  \bibfield  {author} {\bibinfo {author} {\bibfnamefont {H.}~\bibnamefont
  {Esbensen}}, \bibinfo {author} {\bibfnamefont {G.~F.}\ \bibnamefont
  {Bertsch}}, \ and\ \bibinfo {author} {\bibfnamefont {K.}~\bibnamefont
  {Hencken}},\ }\href {\doibase 10.1103/PhysRevC.56.3054} {\bibfield  {journal}
  {\bibinfo  {journal} {Phys. Rev. C}\ }\textbf {\bibinfo {volume} {56}},\
  \bibinfo {pages} {3054} (\bibinfo {year} {1997})}\BibitemShut {NoStop}%
\bibitem [{\citenamefont {Fortunato}\ \emph {et~al.}(2014)\citenamefont
  {Fortunato}, \citenamefont {Chatterjee}, \citenamefont {Singh},\ and\
  \citenamefont {Vitturi}}]{2014Lorenzo}%
  \BibitemOpen
  \bibfield  {author} {\bibinfo {author} {\bibfnamefont {L.}~\bibnamefont
  {Fortunato}}, \bibinfo {author} {\bibfnamefont {R.}~\bibnamefont
  {Chatterjee}}, \bibinfo {author} {\bibfnamefont {J.}~\bibnamefont {Singh}}, \
  and\ \bibinfo {author} {\bibfnamefont {A.}~\bibnamefont {Vitturi}},\ }\href
  {\doibase 10.1103/PhysRevC.90.064301} {\bibfield  {journal} {\bibinfo
  {journal} {Phys. Rev. C}\ }\textbf {\bibinfo {volume} {90}},\ \bibinfo
  {pages} {064301} (\bibinfo {year} {2014})}\BibitemShut {NoStop}%
\bibitem [{\citenamefont {Richter}\ \emph {et~al.}(1990)\citenamefont
  {Richter}, \citenamefont {Weiss}, \citenamefont {Ha\"usser},\ and\
  \citenamefont {Brown}}]{1990Richter}%
  \BibitemOpen
  \bibfield  {author} {\bibinfo {author} {\bibfnamefont {A.}~\bibnamefont
  {Richter}}, \bibinfo {author} {\bibfnamefont {A.}~\bibnamefont {Weiss}},
  \bibinfo {author} {\bibfnamefont {O.}~\bibnamefont {Ha\"usser}}, \ and\
  \bibinfo {author} {\bibfnamefont {B.~A.}\ \bibnamefont {Brown}},\ }\href
  {\doibase 10.1103/PhysRevLett.65.2519} {\bibfield  {journal} {\bibinfo
  {journal} {Phys. Rev. Lett.}\ }\textbf {\bibinfo {volume} {65}},\ \bibinfo
  {pages} {2519} (\bibinfo {year} {1990})}\BibitemShut {NoStop}%
\bibitem [{\citenamefont {Marcucci}\ \emph {et~al.}(2008)\citenamefont
  {Marcucci}, \citenamefont {Pervin}, \citenamefont {Pieper}, \citenamefont
  {Schiavilla},\ and\ \citenamefont {Wiringa}}]{2008Marcucci}%
  \BibitemOpen
  \bibfield  {author} {\bibinfo {author} {\bibfnamefont {L.~E.}\ \bibnamefont
  {Marcucci}}, \bibinfo {author} {\bibfnamefont {M.}~\bibnamefont {Pervin}},
  \bibinfo {author} {\bibfnamefont {S.~C.}\ \bibnamefont {Pieper}}, \bibinfo
  {author} {\bibfnamefont {R.}~\bibnamefont {Schiavilla}}, \ and\ \bibinfo
  {author} {\bibfnamefont {R.~B.}\ \bibnamefont {Wiringa}},\ }\href {\doibase
  10.1103/PhysRevC.78.065501} {\bibfield  {journal} {\bibinfo  {journal} {Phys.
  Rev. C}\ }\textbf {\bibinfo {volume} {78}},\ \bibinfo {pages} {065501}
  (\bibinfo {year} {2008})}\BibitemShut {NoStop}%
\bibitem [{\citenamefont {Moraghe}\ \emph {et~al.}(1994)\citenamefont
  {Moraghe}, \citenamefont {Amaro}, \citenamefont {García-Recio},\ and\
  \citenamefont {Lallena}}]{1994Moraghe}%
  \BibitemOpen
  \bibfield  {author} {\bibinfo {author} {\bibfnamefont {S.}~\bibnamefont
  {Moraghe}}, \bibinfo {author} {\bibfnamefont {J.}~\bibnamefont {Amaro}},
  \bibinfo {author} {\bibfnamefont {C.}~\bibnamefont {García-Recio}}, \ and\
  \bibinfo {author} {\bibfnamefont {A.}~\bibnamefont {Lallena}},\ }\href
  {\doibase https://doi.org/10.1016/0375-9474(94)90744-7} {\bibfield  {journal}
  {\bibinfo  {journal} {Nuclear Physics A}\ }\textbf {\bibinfo {volume}
  {576}},\ \bibinfo {pages} {553 } (\bibinfo {year} {1994})}\BibitemShut
  {NoStop}%
\bibitem [{\citenamefont {Edmonds}(1960)}]{60Edm}%
  \BibitemOpen
  \bibfield  {author} {\bibinfo {author} {\bibfnamefont {A.~R.}\ \bibnamefont
  {Edmonds}},\ }\href@noop {} {\emph {\bibinfo {title} {Angular Momentum in
  Quantum Mechanics}}},\ Princeton Landmarks in Physics\ (\bibinfo  {publisher}
  {Princeton University Press},\ \bibinfo {address} {Princeton, USA},\ \bibinfo
  {year} {1960})\BibitemShut {NoStop}%
\bibitem [{NND()}]{NNDCHP}%
  \BibitemOpen
  \href {http://www.nndc.bnl.gov/chart/} {}\bibinfo {note} {``Chart of
  Nuclides'', National Nuclear Data Center (NNDC);
  http://www.nndc.bnl.gov/chart/}\BibitemShut {NoStop}%
\bibitem [{\citenamefont {Thompson}\ \emph {et~al.}(1977)\citenamefont
  {Thompson}, \citenamefont {Lemere},\ and\ \citenamefont {Tang}}]{77Thom}%
  \BibitemOpen
  \bibfield  {author} {\bibinfo {author} {\bibfnamefont {D.}~\bibnamefont
  {Thompson}}, \bibinfo {author} {\bibfnamefont {M.}~\bibnamefont {Lemere}}, \
  and\ \bibinfo {author} {\bibfnamefont {Y.}~\bibnamefont {Tang}},\ }\href
  {\doibase http://dx.doi.org/10.1016/0375-9474(77)90007-0} {\bibfield
  {journal} {\bibinfo  {journal} {Nuclear Physics A}\ }\textbf {\bibinfo
  {volume} {286}},\ \bibinfo {pages} {53 } (\bibinfo {year}
  {1977})}\BibitemShut {NoStop}%
\bibitem [{\citenamefont {Myo}\ \emph {et~al.}(2001)\citenamefont {Myo},
  \citenamefont {Kat\ifmmode~\bar{o}\else \={o}\fi{}}, \citenamefont {Aoyama},\
  and\ \citenamefont {Ikeda}}]{01Myo}%
  \BibitemOpen
  \bibfield  {author} {\bibinfo {author} {\bibfnamefont {T.}~\bibnamefont
  {Myo}}, \bibinfo {author} {\bibfnamefont {K.}~\bibnamefont
  {Kat\ifmmode~\bar{o}\else \={o}\fi{}}}, \bibinfo {author} {\bibfnamefont
  {S.}~\bibnamefont {Aoyama}}, \ and\ \bibinfo {author} {\bibfnamefont
  {K.}~\bibnamefont {Ikeda}},\ }\href {\doibase 10.1103/PhysRevC.63.054313}
  {\bibfield  {journal} {\bibinfo  {journal} {Phys. Rev. C}\ }\textbf {\bibinfo
  {volume} {63}},\ \bibinfo {pages} {054313} (\bibinfo {year}
  {2001})}\BibitemShut {NoStop}%
\bibitem [{\citenamefont {Suzuki}\ \emph {et~al.}(2004)\citenamefont {Suzuki},
  \citenamefont {Matsumura},\ and\ \citenamefont {Abu-Ibrahim}}]{04Suzu}%
  \BibitemOpen
  \bibfield  {author} {\bibinfo {author} {\bibfnamefont {Y.}~\bibnamefont
  {Suzuki}}, \bibinfo {author} {\bibfnamefont {H.}~\bibnamefont {Matsumura}}, \
  and\ \bibinfo {author} {\bibfnamefont {B.}~\bibnamefont {Abu-Ibrahim}},\
  }\href {\doibase 10.1103/PhysRevC.70.051302} {\bibfield  {journal} {\bibinfo
  {journal} {Phys. Rev. C}\ }\textbf {\bibinfo {volume} {70}},\ \bibinfo
  {pages} {051302(R)} (\bibinfo {year} {2004})}\BibitemShut {NoStop}%
\bibitem [{\citenamefont {Hagino}\ and\ \citenamefont
  {Sagawa}(2007)}]{07Hagi_03}%
  \BibitemOpen
  \bibfield  {author} {\bibinfo {author} {\bibfnamefont {K.}~\bibnamefont
  {Hagino}}\ and\ \bibinfo {author} {\bibfnamefont {H.}~\bibnamefont
  {Sagawa}},\ }\href {\doibase 10.1103/PhysRevC.75.021301} {\bibfield
  {journal} {\bibinfo  {journal} {Phys. Rev. C}\ }\textbf {\bibinfo {volume}
  {75}},\ \bibinfo {pages} {021301(R)} (\bibinfo {year} {2007})}\BibitemShut
  {NoStop}%
\bibitem [{\citenamefont {Myo}\ \emph {et~al.}(2014)\citenamefont {Myo},
  \citenamefont {Kikuchi}, \citenamefont {Masui},\ and\ \citenamefont
  {Kato}}]{14Myo_Rev}%
  \BibitemOpen
  \bibfield  {author} {\bibinfo {author} {\bibfnamefont {T.}~\bibnamefont
  {Myo}}, \bibinfo {author} {\bibfnamefont {Y.}~\bibnamefont {Kikuchi}},
  \bibinfo {author} {\bibfnamefont {H.}~\bibnamefont {Masui}}, \ and\ \bibinfo
  {author} {\bibfnamefont {K.}~\bibnamefont {Kato}},\ }\href {\doibase
  https://doi.org/10.1016/j.ppnp.2014.08.001} {\bibfield  {journal} {\bibinfo
  {journal} {Progress in Particle and Nuclear Physics}\ }\textbf {\bibinfo
  {volume} {79}},\ \bibinfo {pages} {1 } (\bibinfo {year} {2014})}\BibitemShut
  {NoStop}%
\bibitem [{\citenamefont {Tanimura}\ \emph {et~al.}(2012)\citenamefont
  {Tanimura}, \citenamefont {Hagino},\ and\ \citenamefont {Sagawa}}]{2012Tani}%
  \BibitemOpen
  \bibfield  {author} {\bibinfo {author} {\bibfnamefont {Y.}~\bibnamefont
  {Tanimura}}, \bibinfo {author} {\bibfnamefont {K.}~\bibnamefont {Hagino}}, \
  and\ \bibinfo {author} {\bibfnamefont {H.}~\bibnamefont {Sagawa}},\ }\href
  {\doibase 10.1103/PhysRevC.86.044331} {\bibfield  {journal} {\bibinfo
  {journal} {Phys. Rev. C}\ }\textbf {\bibinfo {volume} {86}},\ \bibinfo
  {pages} {044331} (\bibinfo {year} {2012})}\BibitemShut {NoStop}%
\end{thebibliography}

%

\end{document}